\documentclass[a4paper,11pt]{article}
\usepackage{latexsym}
\usepackage{amsmath,amssymb,mathrsfs,pictex} 
\usepackage{amsmath,amsthm,amsfonts,amscd,eucal,bm}
\usepackage{graphicx}
\usepackage{epsfig}
\usepackage{epstopdf} 
\usepackage{cancel}
\usepackage{xcolor}

\numberwithin{equation}{section}

\def\cf{{\mathcal F}}

\def\ch{{\mathcal H}}

\def\bc{{\mathbb C}}
\def\bd{{\mathbb D}}

\def\bn{{\mathbb N}}

\def\br{{\mathbb R}}
\def\bs{{\mathbb S}}
\def\bt{{\mathbb T}}

\def\bz{{\mathbb Z}}

\def\a{\alpha}
\def\b{\beta}
\def\g{\gamma}        \def\G{\Gamma}
\def\d{\delta}        \def\D{\Delta}
\def\eps{\varepsilon} 
     
\def\z{\zeta}
    \def\Th{\Theta}

\def\l{\lambda}       \def\La{\Lambda}
\def\m{\mu}
\def\n{\nu}

\def\r{\rho}
\def\s{\sigma}        \def\S{\Sigma}

        \def\Om{\Omega}

\def\bdk{\boldsymbol{k}}
\def\bdxi{\boldsymbol{\xi}}
\def\bdx{\boldsymbol{x}}

\newtheorem{Thm}{Theorem}[section]

\newtheorem{Prop}[Thm]{Proposition}
\newtheorem{Lemma}[Thm]{Lemma}

\theoremstyle{definition}

\newtheorem{Rem}[Thm]{Remark} 
\theoremstyle{remark}

\def\di{\mathrm{d}}

\newcommand{\tr}{\textrm{Tr}}

\newcommand{\ty}[1]{\mathop{\rm {#1}}}

\newcommand{\nn}{\nonumber}

\title{High-energy asymptotics for the spectral actions of free particles in thermal equilibrium}

\author{{\sc Fabio Ciolli${}^a$,  Francesco Fidaleo}${}^b$\\[12pt]
\small
${}^a$Dipartimento di Matematica e Informatica, Universit{\`a}
della Calabria \\
\small Via Pietro Bucci - Cubo 30B, 87036 Rende (CS), Italy \\[5pt] 
\small
${}^b$Dipartimento di Matematica, Universit{\`a} di Roma Tor Vergata \\
\small Via della Ricerca Scientifica 1, 00133 Roma, Italy \\
} 

\date{}
\begin{document}

\maketitle

\begin{abstract} We provide the asymptotics for high energy regime of the spectral action associated with the extensive quantity describing the average number of  particles of a gas of free particles in thermal equilibrium. We analyse the case of massless particles (the massive non relativistic case being easier and well known) living in a finite volume manifold of arbitrary dimension, and subjected to periodic boundary conditions (i.e. the $d$-dimensional torus), obeying Bose, Fermi and Boltzmann statistics. We also treat in full generality the case of relativistic massive particles, which is more difficult and presents the appearance of a logarithmic term when the dimension of the underlying manifold is odd. The situation relative to
massless and relativistic massive particles living on the 3-sphere is also studied, showing that, even in this case the asymptotics relative to the massive relativistic particles presents an analogous logarithmic term.

\end{abstract}
\smallskip\small{
}

\section{Introduction}

Noncommutative geometry was introduced by Alain Connes in the late 70s and, in the last decades, it 
has grown considerably for many potential applications. The reader is referred to the monograph \cite{Co} for a quite complete and understandable treatise on this topic.
Among the applications to several aspects of mathematics and physics, we would like to mention
the attempt to solve the, still open, long-standing problem of the Riemann 
hypothesis on the location of the zeros of the Riemann zeta function
(e.g.\ \cite{C1}). Concerning the application in quantum field theory, noncommutative geometry might have a role in providing a model, still not available, 
which 
unifies the gravitation with the other three remaining elementary interactions known in nature, see e.g.\ \cite{CC}. 

One of the main ingredients arising in noncommutative geometry is the so-called {\it Dirac operator}. Such an operator is nothing but a kind of square root of the Laplace operator on spin manifolds, up to the
curvature term in the Lichnerowicz formula. In Connes' formulation of noncommutative geometry, the Dirac operator is supposed to encode many of the main properties of the associated, commutative or non commutative, underlying manifold. 
Therefore, the investigation of the spectral properties of the Dirac operator is of crucial relevance: it enters within the definition of the so-called {\it spectral triples}, see e.g. \cite{Co, CM} and the references cited therein. Very recently, 
examples based on the noncommutative 2-tori were constructed and investigated in \cite{FS, CF0}. In such a situation, the involved Dirac operator is suitably deformed by the use of the Tomita modular operator, the latter being unbounded and not ``inner'' in all cases studied there. It is then possible to explicitly construct and study models based on type III representations for which such Tomita modular data play a crucial role.

Another important object, the so-called {\it spectral action},
was first introduced in \cite{CC} with the aim to provide a set of consistent equations in which all fundamental interactions in nature are unified. Also in this case, the Dirac operator $D$ plays a crucial role because the spectral action is supposed to have the form $S(D,\l)=\tr f(|D|/\l)$. Here the function $f$ is associated with an extensive quantity, and $\l$ is a natural cut-off, determined by the properties of the underlying physical system, and describing its behaviour at high energies. As explained in several works (e.g. \cite{CC, CC1, EGV, EI}), the asymptotic expansion of the heat kernel associated with the (powers of the) Dirac operator $D$, and the spectral action $S(D,\l)$, with respect to such a cut-off $\l\uparrow\infty$, is an important tool in many approaches to spectral geometry. 

It should be pointed out that concrete and significant models on which fruitful and exhaustive computations can be carried out are still not completely defined and studied. However, among those we can mention the theories associated with gas of free particles, known in quantum field theory as {\it free fields}. The investigation of the asymptotics of the spectral action associated with extensive quantities, like the entropy, started in \cite{DKS} relatively to integer spin particles (i.e. Bose fields).

In the previous paper \cite{CiF} by the authors, the systematic study of the asymptotics of such extensive quantities has been carried out, extending the study to the so-called $q$-particles (i.e. {\it quons}, see \cite{CF}). 
Even if in principle it is possible to investigate the asymptotics associated with any extensive quantity like the entropy, the analysis was restricted to the extensive quantity describing the ``average number of particles" of such systems made of general free $q$-particles in thermal equilibrium. The asymptotic expansion was performed with respect to the natural cut-off associated with $\l=1/\b=k_{\rm B}T$, $T$ being the absolute temperature and $k_{\rm B}=1.380649\times10^{-23}\,
J K^{-1}$ the Boltzmann constant, or its square root $1/\sqrt{\b}$ in the case of massive non relativistic particles. 
In \cite{CiF}, some cases with connected questions were left open. Among those, we mention the precise meaning of the expansion in Section 6 of \cite{CiF} relative to the Fermi-like particles $q\in[-1,0)$. The case relative to Bose-like particles $q\in(0,1]$ was left open. Another interesting situation, only touched upon in that paper, was that relative to massive relativistic particles. Also, most of the new and significant results in \cite{CiF} have been obtained when the geometrical dimension of the underlying manifold is $d=1$.

In the present paper, we complete the analysis in \cite{CiF}, dealing with particles living on manifolds of finite volume of arbitrary dimension with periodic boundary conditions, that is on the $d$-dimensional tori $\bt^d$. We limit the analysis to the spectral action associated with the number function and, to simplify without losing generality, we reduce the matter
to convenient values of all physical and geometrical involved quantities. We answer all questions left open in \cite{CiF} providing a precise meaning to the asymptotic expansions in Propositions 6.1 and 7.2 there, extending the computations to the Bose case, and thus confirming the conjecture 
that the Bose case behaves exactly like the Fermi case. 
We also exhaust the computations relative to the complete asymptotics concerning massive relativistic particles in all physically interesting cases relative to Boltzmann, Bose and Fermi situations, 
these being much more difficult and interesting. For example, we find a logarithmic term proportional to $\beta\ln\beta$ for $\b\downarrow0$, and explain why it appears only in odd
dimensions of the underlying manifold. The appearance of similar terms in high-temperature expansions for massive
fields is already known in thermal field theory, see e.g. \cite{DoJa}.

We carry out in full generality the analysis relative to the 3-sphere for all relevant cases: massless, and massive relativistic situations, concerning Boltzmann, Bose and Fermi models. After outlining the simpler case of even-dimensional spheres, we show that
the $\beta\ln\beta$ term also appears in the 3-sphere model, inferring that 
this logarithmic term indeed arises only in odd dimension as for previous models.

\section{Preliminaries}

\subsection{Basic facts}
\label{pecoj}

We denote by ${\bf v}=(v_1,\dots,v_d)$ any vector in $\br^d$. We also include the case when ${\bf v}\in\bz^d\subset\br^d$. For 
two such vectors ${\bf v}, {\bf w}$, their inner product is
$$
{\bf v}\cdot {\bf w}:=\sum_{i=1}^d v_iw_i\,.
$$
Therefore
$$
v:=\|v\|=\sqrt{{\bf v}\cdot {\bf v}}=\sqrt{\sum_{i=1}^d v_i^2}\,.
$$

We fix some point in the real line which can be taken as the origin without losing generality. For some $\d>0$, let $f:(-\d,\d)\to\br$ be a real-valued function defined in a neighbourhood of $0$. We say that $f$ is {\it infinitely often differentiable at $0$}, and write 
$f\in C^\infty(0)$ with a slight abuse of notation, if there exists a sequence of strictly positive numbers $0<\d_n\leq\d$, $n\in\bn$, such that $f$ admits the $n$-th derivative $f^{(n)}:(-\d_n,\d_n)\to\br$ for each $n\in\bn$. 
Recall that, by Peano theorem, each $f\in C^\infty(0)$ admits the Maclaurin expansion
\begin{equation}
\label{mclau1}
f(x)=\sum_{k=0}^n\frac{f^{(k)}(0)}{k!}x^k+R_n(x)\,,\quad x\in(-\d_n,\d_n)\,,n\in\bn\,,
\end{equation}
and, for the remainder, $R_n(x)=o(x^n)$. In this case, we write
\begin{equation}
\label{mclau2}
f(x)\sim\sum_{n=0}^{+\infty}\frac{f^{(n)}(0)}{n!}x^n\,\,\text{for}\,\,x\to0\,.
\end{equation}

At this stage, it is natural to discuss the following questions connected with the converse of Peano theorem. Indeed, consider a sequence
$(a_k)_{k\in\bn}\subset\br$ and a function $f$  
defined in an open neighbourhood 
of 0. Suppose that
\begin{equation}
\label{convpea1}
f(x)=\sum_{k=0}^n a_k x^k+o(x^{n})\,,\quad n\in\bn\,.
\end{equation}
We list some situations which can appear:
\begin{itemize}
\item[$1^o$] the power series $\sum_{k=0}^{+\infty} a_k x^k$ has nonzero convergence radius, $r>0$;
\item[$2^o$] the power series $\sum_{k=0}^{+\infty} a_k x^k$ has zero convergence radius.
\end{itemize}
The $1^o$ case splits into two sub-cases:
\begin{itemize}
\item[$1^o$-a] the power series $\sum_{k=0}^{+\infty} a_k x^k$ coincides with $f$ where it converges and, consequently, $f$ extends to the holomorphic function $\sum_{k=0}^{+\infty} a_k z^k$, in the disc $|z|<r$ in the complex plane;
\item[$1^o$-b] the power series $\sum_{k=0}^{+\infty} a_k x^k$ does not coincide with $f$ where it converges.
\end{itemize}
Among the situations described in $1^o$-b and in the $2^o$ case, we find:
$1^o$-b-(i) and $2^o$-(i), $f\in C^\infty(0)$; $1^o$-b-(ii) and 
$2^o$-(ii), $f\notin C^\infty(0)$.

Notice that, by Peano theorem, in the cases $1^o$-a, $1^o$-b-(i) and $2^o$-(i), we immediately conclude that, for each $k\in\bn$,
$a_k=\frac{f^{(k)}(0)}{k!}$,
and \eqref{mclau1} holds true. 
For $f$ in the classes $1^o$-b-(i) and $2^o$-(i) (for the holomorphic case $1^o$-a, we directly use the equality), we use the notation in 
\eqref{mclau2}.

We now provide examples relative to all pathological (i.e. non holomorphic) cases listed above.
Define
\begin{equation}
\label{convpea}
f(x):=
\begin{cases}
e^{-1/x^2}\,,\,\, x\neq0\,,\\
0\,,\,\,\, x=0\,,
\end{cases}\quad g(x):=\int_0^{+\infty}\frac{e^{-t}}{1+x^2t}\di t\,.
\end{equation}
It is easy to check that $f(x)$ provides an example of $1^o$-b-(i), whereas $f(x)\sin(1/f(x))$, extended to 0 at 0, instead provides an example of
$1^o$-b-(ii).\footnote{We are indebted to the referee for this pathological example $f(x)\sin(1/f(x))$.}
An example of $2^o$-(ii) is easily given by $f(x)\sin(1/f(x))+g(x)$, whereas $g(x)$ alone provides an example for $2^o$-(i).
Therefore, the function $f(x)\sin(1/f(x))$ provides a counterexample to the converse of Peano theorem in the following sense: if we have a function satisfying \eqref{convpea1}, it is not necessarily of the class $C^\infty(0)$.\footnote{In the case of $f(x)\sin(1/f(x))$, $a_k=0$ in 
 \eqref{convpea1}, identically.}

A quantity $F=o\big((x-x_o)^\infty\big)$ as $x\to x_o$, simply means
\begin{equation}
\label{mclau223}
\lim_{x\to x_o}F(x)/(x-x_o)^n=0\,,\quad \text{for all}\,\,n\in\bn\,.
\end{equation}
Notice that, in all situations treated in the present paper, $x_o=0$ and $x=\beta$, where $\b$ is the inverse temperature of a system in thermal equilibrium.

For a function $f\in C^\infty(0)$, a quite delicate question is to decide when, and in which sense, in \eqref{mclau2} we should add a further term as
\begin{equation}
\label{mclau222}
f(x)\sim\sum_{n=0}^{+\infty}\frac{f^{(n)}(0)}{n!}x^n+o(x^\infty)\,\,\text{for}\,\,x\to0\,,
\end{equation}
see \eqref{mclau223} for the definition of $o(x^\infty)$. If the series in \eqref{mclau222} (or, equivalently, in \eqref{mclau2}) has nonzero convergence radius, such a term can appear only when $f$ is not holomorphic in a neighbourhood of 0, that is $1^o$-b-(i) above. This is precisely the case described in \eqref{terda}. Therefore, we reduce the matter to the remaining case $2^o$-(i). 

If we consider the functions
$g(x)$ and $f(x)+g(x)$, $f,g$ given in \eqref{convpea}, both have the (divergent) Euler series $\sum(-1)^n n! u^n$, $u:=x^2$, as Maclaurin expansion, but the inversion of the Borel summation process only produces $g(x)$. In this sense, we can say that the term $o(x^\infty)$ appears in the latter but not in the former. A similar situation would appear below concerning the asymptotic expansions relative to all Bose and Fermi cases. In such situations, we should consider generalised summation processes for the involved divergent series, according to the growth of the coefficients (e.g. \cite{Cos}), and then show that we re-obtain the original function. In this sense, we might infer that the term $o(\b^\infty)$ would never appear in our computations.

We decided not to pursue further this fundamental and fascinating aspect of asymptotic analysis, and thus we do not provide the proof of questions relative to the inversion of the summation processes outlined above, and refer the reader to the existing literature. Therefore, for the asymptotic expansions involving non convergent series, we always use the notation in \eqref{mclau2} without further comments.

\subsection{The spectral action for free particles}

The spectral action for a gas of free particles in thermal equilibrium can be naturally associated with thermodynamic potentials, which are extensive quantities like the entropy (e.g. \cite{CCS}), the mean energy
or the number function (e.g. \cite{CiF}). They can be recovered by the {\it grand partition function}, see e.g. \cite{Hu}. 

Indeed, we consider a system whose Hamiltonian $H$ is a self-adjoint positive (i.e.\ $\s(H)\subset[0,+\infty)$) operator with compact resolvent, acting on a separable Hilbert space $\ch$, called the {\it one-particle space}.
In such a situation, the spectrum $\s(H)$ is made of isolated points, accumulating at $+\infty$ if $\ch$ is infinite dimensional. The multiplicity 
$g(\eps)$ of each eigenvalue $\eps\in\s(H)$ is necessarily finite. We then have 
$I_\ch=\sum_{\eps\in\s(H)}P_{\eps}$, and
$$
H=\sum_{\eps\in\s(H)}\eps P_{\eps},\,\,\text{and}\,\,g(\eps):=\text{dim}\big(\text{Ran}(P_{\eps})\big)<\infty\,.
$$

We also suppose that, at any inverse temperature $\b:=1/k_B T$ ($k_B$ being the {\it Boltzmann constant}), $e^{-\b H}$ is trace-class for each $\b>0$, and define the {\it partition function}
$\z:=\ty{Tr}\!\big(e^{-\b H}\big)$. 

It is well known that the fundamental operators of systems describing noninteracting particles, made of annihilators and creators acting on the corresponding Fock spaces, satisfy the commutation relations
\begin{equation*}
{\bf a}_q(f){\bf a}_q^\dagger(g)-q{\bf a}_q^\dagger(g){\bf a}_q(f)=\langle g,f\rangle I_\ch\,, \quad f,g\in\ch\,,
\end{equation*}
where $q=\pm1$ correspond to Bose and Fermi particles, respectively, see e.g. \cite{BR}. It is possible to extend the analysis to the cases $q\in[-1,1]$ relative to the  
so-called $q$-{\it particles} (e.g. \cite{BoKuSp}), usually called {\it quons} for the cases $q\neq\pm1,0$. The case $q=0$ also has fundamental importance because it corresponds to classical particles, that is those obeying Boltzmann statistics.

Notice that, to recover the right grand partition function for $q\in(-1,1)$ is a very delicate question, solved for the Boltzmann case $q=0$ by correcting the involved formulas by the Gibbs factors $1/n!$ which take into account the fundamental principle in quantum mechanics of indistinguishability of particles. The general solution for all $q\in[-1,1]$, including the known ones for Bose, Fermi and Boltzmann particles, is provided in \cite{CF}.

For the convenience of the reader, we report here the formula for the Grand partition function $Z_q$, recalling that we will limit our analysis to the physically relevant cases $q=0,\pm1$:
\begin{equation*}
\begin{split}
Z(\b,z):=Z_q=e^{\displaystyle -\frac{\displaystyle \tr\ln\left(I-zqe^{\displaystyle -\b H}\right)}{\displaystyle q}}\,,&\quad 
\left\{\begin{array}{ll}
0<z<\frac{\displaystyle e^{\displaystyle \b\min\s(H)}}{\displaystyle q}\,\,& 0<q\leq1\,, \\
{}&\\
z>0\,\,& -1\leq q<0\,.
\end{array}\right.\\
Z_0=e^{z\tr\big(e^{-\b H}\big)}\,,&\quad z>0\,.
\end{split}
\end{equation*}
Here $z:=e^{\b\m}$ is the {\it activity} (or {\it fugacity}) corresponding to the {\it chemical potential} $\m$. The reader is referred to \cite{Hu} for the main notations and basic results in statistical mechanics.

We now note that any extensive quantity which can provide a suitable spectral action
like the entropy and the Landau potential, can be recovered by using the grand partition function. The simplest extensive quantity is the {\it number function}, describing the average number of particles in thermal equilibrium, which is obtained as
\begin{equation}
\label{occqqqq3}
N(\b,z)=z\frac{\partial\ln Z(\b,z)}{\partial z}={\rm Tr}\big(z^{-1}e^{\b H}-q\big)^{-1}\,.
\end{equation}
As in \cite{CiF}, we limit our analysis to the spectral action associated with such a number function.

For such free models, the Hamiltonian is a suitable function of the Dirac operator $D$, by thinking of $D$ as proportional to ``the square root" of the Laplacian. Indeed,
\begin{equation*}
H(|D|)\propto\left\{\begin{array}{ll}
                    \!\!\!\!\!\!\! &|D|\,\,\text{for massless case}\,, \\[1ex]
               	\!\!\!\!\!\!\!&D^2\,\,\text{for massive case}\,. \\[1ex]
                    \end{array}
                    \right.
\end{equation*}

We note that, the massless case pertains to relativistic situations (i.e.\ photons or, even, gravitons), as well as to nonrelativistic situations such as phonons. In these cases, the constant $c$ appearing in \eqref{sacc0000} corresponds to the speed of light or the speed of sound in the considered medium, respectively.
Here instead, the massive case is always referred to the non relativistic situation. The relativistic situation treated in the 2nd half of the present paper is obviously referred to the massive case.

As explained in \cite{DKS} (see \cite{CC} for the original definition), a spectral action is a function of the Dirac operator which has the form
\begin{equation}
\label{sacc}
S=S(f,D,\l):=\tr\big(f(|D|/\l)\big)\,,
\end{equation}
when the r.h.s.\ is meaningful and, for the cut-off,
\begin{equation*}
\l=\left\{\begin{array}{ll}
                    \!\!\!\!\!\!\! &1/\b\,\,\text{for the massless case}\,, \\[1ex]
               	\!\!\!\!\!\!\!&1/\sqrt{\b}\,\,\text{for the massive non relativistic case}\,, \\[1ex]
                    \end{array}
                    \right.
\end{equation*}
with $\b$ (proportional to) the inverse temperature.\footnote{We have $\l=1/\b$ also for the massive relativistic case, see Section \ref{mhdrcx}.}
Therefore, in our situation here, 
\begin{equation}
\label{occiiii}
f=\big(z^{-1}e^{\b H(|D|)}-q\big)^{-1}\,,
\end{equation}
see \eqref{occqqqq3}.

We end by noticing that, first \eqref{occiiii} reduces to  $e^{-\b H(|D|)}$ (neglecting the multiplicative factor $z$) when $q=0$, and thus the involved spectral action is proportional to the partition function
$$
\z(\beta)={\rm Tr}\big(e^{-\b H(|D|)}\big)
$$
in the Boltzmann case. Second, the spectral action associated with \eqref{occiiii} is meaningful (i.e.\ the involved operators are  trace class), see e.g.\ \cite{CiF}, Proposition 3.1.

\subsection{The asymptotics to high energies of the spectral action}

We recall some results, already known (e.g. \cite{CC1, CiF, DKS, EI}),
for the asymptotics at high energies, that is when the temperature of the system is going to $+\infty$ or, equivalently, $\b\downarrow0$.

We first consider the spectral actions, after passing to the continuum, for infinitely extended systems living on $\br^d$, for which the involved Dirac operator, or equivalently the Laplacian, has continuum spectrum. 
In such a situation,
\begin{equation}
\label{sacc0000}
H=\left\{\begin{array}{ll}
                    \!\!\!\!\!\!\! &cp\,\,\text{for massless case}\,, \\[1ex]
               	\!\!\!\!\!\!\!&\frac{p^2}{2m}\,\,\text{for massive case}\,. \\[1ex]
                    \end{array}
                    \right.
\end{equation}
Here the Dirac operator (after the Fourier transform) is the momentum vector ${\bf p}$ whose modulus is 
$$
p:=\sqrt{{\bf p}\cdot{\bf p}}=\sqrt{\sum_{i=1}^d p_i^2}\,,
$$
$c = 299792458\,m\,s^{-1}$ is the speed of light, and $m$ is the mass of the involved particles. If, for example, the massless case concerns the phonons, $c$ should be considered as the speed of sound which, obviously, depends on the involved medium.

It is customary (e.g.\ \cite{Hu}) to perform the passage to the continuum as follows:
for a gas with $N$ particles in the volume $V_d$ in the $d$-dimensional space, we simply make the replacements
\begin{equation*}
\sum_{\eps\in\s(H)}g(\eps)=\sum_{\{n(\eps)\,;\,\eps\in\s(H)\}}\rightarrow
V_d\int\frac{\di^d{\bf p}}{h^d}\,,
\end{equation*}
$h = 6.62607015\times10^{-34}\,m^2\,kg\,s^{-1}$ being the Planck constant, and $V_d$ the $d$-dimensional volume of the physical system under consideration.

For the convenience of the reader, we collect some results contained in \cite{CiF} relative to the asymptotics of the spectral action in this situation. In so doing, we denote by $\Om_d$, $d\geq1$, the $d$-dimensional ``solid" angle, that is the measure of the $(d-1)$-hypersurface of the hypersphere of radius 1 in $\br^d$. For the involved spectral action $S$ in \eqref{sacc} associated with the number function \eqref{occiiii}, we simply write $S_{\rm Number}(\b)$.
\medskip

\noindent
\textbf{Massless case for Fermi particles $q=-1$}:
$$
S_{\rm Number}(\b)=\left(\frac{V_d\Om_d}{h^{d-1}c^d}\int_0^{+\infty}\di y\frac{y^{d-1}}{z^{-1}e^y+1}\right)\frac1{\b^d}\,.
$$
\medskip

\noindent
\textbf{Massless case for the Boltzmann case $q=0$}: 
$$
S_{\rm Number}(\b)=\left(\frac{zV_d\Om_d}{h^{d-1}c^d}\int_0^{+\infty}\di y\,y^{d-1}e^{-y}\right)\frac1{\b^d}\,.
$$
\noindent
\textbf{Massive case for Fermi particles $q=-1$}:
$$
S_{\rm Number}(\b)=\left(\frac{V_d\Om_d(2m)^{d/2}}{h^{d-1}}\int_0^{+\infty}\di y\frac{y^{d-1}}{z^{-1}e^{y^2}+1}\right)\frac1{\b^{d/2}}\,.
$$
\medskip

\noindent
\textbf{Massive case for the Boltzmann case $q=0$}:
$$
S_{\rm Number}(\b)=\left(\frac{zV_d\Om_d(2m)^{d/2}}{h^{d-1}}\int_0^{+\infty}\di y\,y^{d-1}e^{-y^2}\right)\frac1{\b^{d/2}}\,.
$$

The Bose case deserves particular attention due to the phenomenon of condensation of particles in the ground state, known as {\it Bose-Einstein condensation}, see \cite{BR, Hu, CF}.
It is well known that the condensation takes place when the {\it critical density} is finite:
\begin{align*}
\r_{\rm c}:=\lim_{z\uparrow 1}(N/V_d)\propto\lim_{z\uparrow1}\int_0^{+\infty}\di y\frac{y^{d-1}}{z^{-1}e^{y^\a}-1}
=\int_0^{+\infty}\di y\frac{y^{d-1}}{e^{y^\a}-1}<+\infty\,.
\end{align*}
Here $
\a=\left\{\begin{array}{ll}
                    \!\!\!\!\!\!\! &1\,\,\text{for massless case}\,, \\[1ex]
               	\!\!\!\!\!\!\!&2\,\,\text{for massive case}\,, \\[1ex]
                    \end{array}
                    \right.$ and the above integral is convergent if and only if $\a-d+1<1$. Therefore, the condensation takes place if and only if  
                    $\left\{\begin{array}{ll}
                    \!\!\!\!\!\!\! &d\geq2\,\,\text{for massless case}\,, \\[1ex]
               	\!\!\!\!\!\!\!&d\geq3\,\,\text{for massive case}\,. \\[1ex]
                    \end{array}
                    \right.$  
It is then customary to separate the contribution corresponding to ground level $\eps=0$ from the remaining ones, see e.g.\ \cite{Hu}, and Section 5 of \cite{CF}. Therefore, we set
$$
a\,\,\left\{\begin{array}{ll}
                    \!\!\!\!\!\!\! &=0\,\,\text{for massless case when $d=1$ and massive case when $d=1,2$}\,, \\[1ex]
               	\!\!\!\!\!\!\!&\geq0\,\,\text{for massless case when $d\geq2$ and massive case when $d\geq3$}\,, \\[1ex]
                    \end{array}
                    \right.
$$
where $a$, describing the strength of the possible amount of condensate, is a dimensionless number. Such an amount of condensate will depend on 
the initial density of the system, as well as on the 
boundary conditions (i.e. Dirichlet or Neumann for example) imposed to the Laplace operator of the finite volume theories approaching to the infinite volume one, 
see e.g. \cite{BR}, Section 5.2.5. Therefore, the number $a$ will take into account the possible presence of the condensate after the thermodynamic limit. Concerning the interpretation of the condensation of photons, we refer the reader to \cite{KSVW}.

For Bose particles we have the following situation. 
\medskip

\noindent
\textbf{Massless case for Bose particles $q=1$}:
\begin{equation}
\label{1asdt1}
S_{\rm Number}(\b)=\left(\frac{V_d\Om_d}{h^{d-1}c^d}\int_0^{+\infty}\di y\frac{y^{d-1}}{z^{-1}e^y-1}\right)\frac1{\b^d}
+a\frac{hz}{1-z}\,.
\end{equation}
\medskip

\noindent
\textbf{Massive case for Bose particles $q=1$}:
\begin{equation}
\label{1asdt0}
S_{\rm Number}(\b)=\left(\frac{V_d\Om_d(2m)^{d/2}}{h^{d-1}}\int_0^{+\infty}\di y\frac{y^{d-1}}{z^{-1}e^{y^2}-1}\right)\frac1{\b^{d/2}}
+a\frac{hz}{1-z}\,.
\end{equation}

It should be noted the following facts.
\begin{itemize}
\item[(i)] From a mathematical point of view (e.g. \cite{BR}), the condensation takes place when the chemical potential $\m\approx 0^-$ or, equivalently,
the activity $z\approx 1^-$. In this case, $a\frac{hz}{1-z}$ in \eqref{1asdt1} and  \eqref{1asdt0} describes the amount of the condensate. It is negligible for $z\approx0$ and cannot be neglected when $z\approx1$, provided $a>0$. Notice that, since the finite volume theories are nothing else than the real description of the infinite volume ones, 
it is customary to add similar terms describing the condensate also in finite volume situations, in order to provide a more realistic description of the model.
\item[(ii)] All cases described above can appear as physically meaningful situations. Indeed, concerning Bose massless particles, we mention
photons and gravitons, provided the latter exist. We also mention the quasi-particles, like phonons and many others, which obey Bose statistics. 

For the Fermi massless case, we cite the massless neutrinos, as a purely theoretic model since the known
families of neutrinos must have a mass. This is due to the quantum phenomenon, called ``oscillation'', when a neutrino changes its ``flavour'' (identity) as it travels through space. 

Massless particles obeying Boltzmann statistics do not exist. However, it should be noted that any system in its dilute regime (i.e. when its density $\r\approx0$ or, equivalently, $z\approx0$) obeys Boltzmann statistics. Therefore, the Boltzmann massless case also has a physical meaning.
\item[(iii)] Apart from these constant terms $a\frac{hz}{1-a}$ describing the condensation phenomena, in the infinite volume theories only the divergent part appears in the asymptotics of the spectral action. It is indeed well known that, for the finite volume theories, a plethora of terms in $\b^j$, 
$-d\leq j<+\infty$ will appear in the massless case. The situation for the massive relativistic case is still more complicated for the appearance of logarithmic terms containing $\b\ln\b$ for $\b\downarrow0$.
We analyse below these relevant situations.
\end{itemize}       
                    
Concerning the finite volume theory, we start by noticing that the momentum operator ${\bf p}$ of a particle living in the manifold $[0,L]^d\sim\bt^d$ (i.e. under periodic conditions), is given by
$$
p_i:=-\imath\hbar\frac{\partial\,\,}{\partial x_i}\,,\quad i=1,2,\dots, d\,,
$$           
acting on $L^2([0,L]^d,\di^d x)$. 
By passing in momentum space through the Fourier transform and using the wave-vector ${\bf k}\in\bz^d$, by imposing periodic boundary conditions, we easily compute
$$
{\bf p}=\frac{h}{L}{\bf k}\,,\quad {\bf k}\in\bz^d\,,
$$
and thus
$$
p^\a=\Big(\frac{h}{L}\big({\bf k} \cdot {\bf k}\big)^{1/2}\Big)^\a\,,\quad {\bf k}\in\bz^d
$$
where, as before, $\a=1$ in the massless case, and $\a=2$ in the massive case.\footnote{The formula for the massive relativistic case is provided in \eqref{mrcz}.}

Concerning the asymptotics, we report the simplest massive case whose proof is given in \cite{EI}, Proposition 2.27 (see also \cite{EGV}, Lemma 2.10).
\medskip

\noindent
\textbf{Massive case, $d$ arbitrary}:
$$
S_{\rm Number}(\b)=\left(\frac{L^d\Om_d(2m)^{d/2}}{h^{d-1}}\int_0^{+\infty}\di y\frac{y^{d-1}}{z^{-1}e^{y^2}-q}\right)\frac1{\b^{d/2}}+o\big(\b^\infty\big)\,,\,\,\,\text{as}\, \,\,\b\downarrow0\,,
$$
where, for the activity, $z>0$ for $q=-1,0$, and $0<z<1$ for $q=1$.

As a pivotal example, we report the situation for $d=1$ and $q=0$, for which the spectral action can be explicitly computed. The massless case is nothing more than an exercise:
\begin{equation*}
\begin{split}
S_{\rm Number}(\b)=&zh\sum_{k\in\bz}e^{-\frac{ch\b}{L}|k|}=zh\frac{e^{\frac{ch\b}{L}}+1}{e^{\frac{ch\b}{L}}-1}\\
=&\frac{2zL}{c}\frac1{\b}+2zh\sum_{n=1}^{+\infty}\Big(\frac{ch}{L}\Big)^{2n-1}\frac{B_{2n}}{(2n)!}\b^{2n-1}\,,
\end{split}
\end{equation*}
where $B_{2n}$ are the Bernoulli numbers.
The massive case can be computed in terms of Jacobi Theta function (e.g.\ \cite{WW}, pag. 464), obtaining
\begin{equation}
\label{terda}
\begin{split}
S_{\rm Number}(\b)=&\sqrt{2m\pi}zL\frac1{\sqrt{\b}}+2\sqrt{2m\pi}zL\frac1{\sqrt{\b}}
\sum_{n=1}^{+\infty}e^{-\frac{2mL^2\pi^2n^2}{h^2\b}}\\
=&\sqrt{2m\pi}zL\frac1{\sqrt{\b}}+o\big(\b^\infty\big)\,,\,\,\,\text{as}\, \b\downarrow0\,.
\end{split}
\end{equation}

From now on, for the sake of simplicity and without losing generality, we reduce all quantities by choosing appropriate values of involved parameters and physical quantities. 
The physically meaningful objects can be easily recovered after elementary rescaling. We disregard the condensation effects for the Bose case which, though, might provide only a constant term proportional to the amount of the condensate. We also disregard multiplicative factors like the elementary quantum of action $h$.
For the computation of the asymptotics in Bose and Fermi case, for the activity in \eqref{occiiii} we put $z=1/2$, which is allowed for all situations treated here. 

We deal with the massless and the massive relativistic cases, where the underlying manifold is either the $d$-dimensional torus or the 3-sphere since, the 1-sphere is nothing else than 1-torus, and all even-dimensional spheres can be easily treated, see e.g. Section \ref{outshj}.

\section{Massless cases}

\subsection{Boltzmann case for tori}
\label{1mic1}

We consider the case with
\begin{equation}
\label{essebe}
S(\b):=\sum_{{\bf k}\in\bz^d}e^{-\beta k}\,,\quad\beta>0\,.
\end{equation}
It is known that $S(\b)=L(\b)+R(\b)$, 
where the leading term (e.g. \cite{CiF}, Section 4) is given by
$$
L(\b)=\frac{2\pi^{d/2}\G(d)}{\G(d/2)}\frac1{\beta^d}\,.
$$
The remainder is then defined by 
\begin{equation}
\label{toty00}
R(\beta) = S(\beta) - L(\beta)\,,\quad \b>0\,.
\end{equation}
By using the Fourier transform of the function $f(\bdx)$ (e.g. \cite{grafakos}, Definition 2.2.8),
$$
\widehat{f}(\bdxi)= \cf\{f\}(\bdxi)= \int_{\mathbb{R}^d} f(\bdx)\ e^{-2\pi\imath\,\bdxi \cdot\bdx}\di^d\bdx\ ,\quad \bdx,\bdxi\in\mathbb{R}^d\,,
$$
this computation can be carried out by using the {\it Poisson summation formula}, e.g.  \cite{grafakos}, Theorem 3.2.8:
\begin{equation*}
\sum_{{\bdk} \in \mathbb{Z}^d} f(\bdk) = \sum_{\bdxi\in \mathbb{Z}^d} \hat{f}(\bdxi)\,,
\end{equation*}
where $\bdx\in\mathbb{Z}^d$ and $\bdxi\in\mathbb{Z}^d$ belong to the lattice and the dual one, respectively, applied to the function  
$f(\bdx)=e^{-\beta x}$.
Since $\b>0$ is kept fixed during the following analysis, we drop such a dependence if it causes no confusion.
\begin{Lemma}
\label{pre ex49}
For the above remainder \eqref{toty00}, we have
\begin{equation}
\label{R(b)}
R(\beta)=2^d\pi^{(d-1)/2}\G((d+1)/2)\sum_{{\bf k}\neq0}\frac{\b}{(\b^2+4\pi^2k^2)^{(d+1)/2}}\,,\quad \b>0\,.
\end{equation}
\end{Lemma}
\begin{proof}
For the Lévy–Smirnov function \cite[Eq. 6]{Penson}, 
\begin{equation*}
g(t) = \frac{\beta}{2\sqrt{\pi} t^{3/2}} e^{-\beta^2/4t}\,,\quad t\in(0,+\infty)\, ,
\end{equation*} 
and its Laplace transform $\mathcal{L}$, we start by noticing that
\begin{equation}
\label{laeq}
\begin{split}
\mathcal{L}\{g\}(x^2)&= \int_0^\infty e^{-x^2t} \left( \frac{\beta}{2\sqrt{\pi} t^{3/2}} e^{-\beta^2/4t} \right) \di t \\
        &= \frac{\beta}{2\sqrt{\pi}} \int_0^\infty t^{-3/2}  e^{-\beta^2/4t}  e^{-x^2t} \di t\, =e^{-\beta x}=f(\bdx)\ .
\end{split}
\end{equation}

Concerning the Fourier transform of $f(\bdx)$, we use the last equalities in \eqref{laeq} and exchange the integration by Tonelli-Fubini theorem, obtaining
\begin{align}
\label{aslio}
\hat{f}(\bdxi)=&\int_{\mathbb{R}^d}\di^d\bdx\,e^{-\beta x} e^{-2\pi\imath\,\bdxi \cdot\bdx}
=\frac{\beta}{2\sqrt{\pi}}\int_{\mathbb{R}^d}\di^d\bdx\,e^{-2\pi\imath\,\bdxi \cdot\bdx}\int_0^{+\infty }\di t\,t^{-3/2}  e^{-\beta^2/4t}  e^{-x^2t}\nn\\
=&\frac{\beta}{2\sqrt{\pi}}\int_0^{+\infty }\di t\,t^{-3/2}  e^{-\beta^2/4t} \int_{\mathbb{R}^d}\di^d\bdx\,
e^{-2\pi\imath\,\bdxi \cdot\bdx}e^{-x^2t}\,.
\end{align}
Now, using the well-known result
\[
\mathcal{F}\big\{e^{-tx^2}\big\}(\bdxi) = (\pi/t)^{d/2} e^{-\pi^2\xi^2/t}
\]
in \eqref{aslio}, we obtain
\begin{equation}
\label{ewbo}
\int_{\mathbb{R}^d} e^{-tx^2} e^{-2\pi i \bdxi \cdot \bdx}\,\di^d\bdx = \left(\frac{\pi}{t}\right)^{d/2} e^{-\frac{\pi^2|\xi|^2}{t}}\,.
\end{equation}
Collecting together \eqref{aslio} and \eqref{ewbo}, we get
\begin{equation}
\label{eq:integral_t}
\begin{split}
\hat{f}(\bdxi) =&\frac{\beta}{2\sqrt{\pi}} \int_0^\infty t^{-3/2} e^{-\beta^2/4t} \left(\frac{\pi}{t}\right)^{d/2} e^{-\frac{\pi^2\xi^2}{t}}\,\di t\\
=&\frac{\beta \pi^{(d-1)/2}}{2} \int_0^\infty t^{-\frac{d+3}{2}} \exp\Big( - \frac{\beta^2 + 4\pi^2\xi^2}{4t} \Big)\,\di t\ .
\end{split}
\end{equation}

To evaluate the integral in \eqref{eq:integral_t}, we perform the elementary change of variables
$u:= \frac{\beta^2 + 4\pi^2\xi^2}{4t}$ and, taking into account the definition of Euler Gamma function $\G$ (e.g. \cite{Con}, Ch. VII. §7),
we obtain
\begin{equation*}
\begin{split}
\hat{f}(\bdxi)=&\frac{\beta \pi^{(d-1)/2}}{2} \int_0^\infty e^{-u} \left( \frac{4u}{\beta^2 + 4\pi^2\xi^2} \right)^{\frac{d+3}{2}} \frac{\beta^2 + 4\pi^2\xi^2}{4u^2}\,\di u\\
=&\frac{2^d \pi^{(d-1)/2}  \beta}{\left(\beta^2 + 4\pi^2\xi^2\right)^{(d+1)/2}}
\int_0^{+\infty}u^{(d-1)/2}e^{-u}\,\di u
=\frac{2^d \pi^{(d-1)/2} \Gamma\left(\frac{d+1}{2}\right) \beta}{\left(\beta^2 + 4\pi^2\xi^2\right)^{(d+1)/2}}\,.
\end{split}
\end{equation*}
Now, since for the leading term, 
\begin{equation}
\label{L beta}
\hat{f}(0) =\ \frac{2^d \pi^{(d-1)/2} \Gamma\left(\frac{d+1}{2}\right)}{\beta^d} 
=\frac{2\pi^{d/2}\G(d)}{\G(d/2)\beta^d}=L(\beta)\,, 
\end{equation}
we obtain the assertion.
\end{proof}

\begin{Prop}
\label{anal}
The function $R(\beta)$ admits an analytic continuation into the disk of radius $2\pi$ in the complex plane around the origin.
\end{Prop}
\begin{proof}
It is enough to consider the functions
$$
f_{\bf k}(z):=\frac1{(z^2+4\pi^2k^2)^{(d+1)/2}}\,,\quad k_i>1,\,\,i=1,\dots,d\,,
$$
according to the convention on the square-root in the complex
plane,
such that for integer $l$, $w^{l+1/2}=w^{l}\sqrt{w}$ and
$$
\sqrt{w}=\sqrt{|w|}e^{\imath\arg(w)/2}\,,|w|\in(0,+\infty),\,\,\arg(w)\in(-\pi,\pi)\,.
$$

After some straightforward computations, for $|z|<2\pi$ we get 
$$
|z^2+(2\pi k)^2|\geq 4\pi^2(k^2-1)\,,
$$
and thus
$$
0\leq|f_{\bf k}(z)|\leq\frac1{\big[4\pi^2(k^2-1)\big]^{(d+1)/2}}\,,\quad k^2>1\,.
$$
By a standard comparison with an integral, where $k\to\r$ and $u=\r^2-1$, it is now easy to deduce that
\begin{align*}
\sum_{k>1}|f_{\bf k}(z)|\leq&\sum_{k>1}\frac1{\big[4\pi^2(k^2-1)\big]^{(d+1)/2}}\sim\int_3^{+\infty}\frac{u^{\frac{d-2}2}}{u^{\frac{d+1}2}}\di u\\
=&
\int_3^{+\infty}\frac{\di u}{u^{3/2}}<+\infty\,.
\end{align*}

After reasoning in the same way for the derivative with respect to $z$ and applying the Weierstrass $M$-test to the series of the $f_{\bf k}$ and their derivatives, we immediately deduce that 
$$
R(\b)=2^d\pi^{(d-1)/2}\G((d+1)/2)\sum_{{\bf k}\neq0}\frac{\b}{(\b^2+4\pi^2k^2)^{(d+1)/2}}
$$
converges totally in each disk $\bd_r:=\{z\in\bc\,;\,|z|\leq r\}$, $r<2\pi$ together with the series of their derivatives. Therefore,
$R(z)$ is differentiable, hence holomorphic, in the open disk $\{z\in\bc\,;\,|z|<2\pi\}$.
\end{proof}

We go ahead with our analysis by expressing $S(\b)$ in a more convenient way. Indeed, by expressing the exponential in terms of the inverse Mellin transform of the Gamma function, known as the Cahen-Mellin integral \cite[Sec. 3.3.1]{PaKa}, as
$$
e^{-x}=\frac1{2\pi\imath}\int^{c+\imath\infty}_{c-\imath\infty}\G(s)x^{-s}\di s\,,\quad x,c>0,
$$
we have
\begin{equation}
\label{toty}
\begin{split}
S(\b)=&1+\sum_{{\bf k}\neq0}\frac1{2\pi\imath}\int^{c+\imath\infty}_{c-\imath\infty}\G(s)\b^{-s}k^{-s}\di s\\
=&1+\frac1{2\pi\imath}\int^{c+\imath\infty}_{c-\imath\infty}\G(s)\b^{-s}Z_d(s/2)\di s\,,\quad c>d\,.
\end{split}
\end{equation}
Here $Z_d(z)$ is the meromorphic function obtained by the analytic continuation of
$$
Z_d(z):=\sum_{{\bf k}\in\bz^d\smallsetminus\{0\}}\frac1{k^{2z}}\,,\quad{\rm Re}(z)>d/2
$$
known as the {\it Epstein Z-function} \cite[Sec. 1.4]{TerrasI}, and the exchange of the sum with the integral in \eqref{toty} is justified by Tonelli-Fubini theorem since
\begin{equation}
\label{toty2}
|\G(c+\imath y)|\sim |y|^{c-1/2}e^{-\pi|y|/2}\,,
\end{equation}
and
$$
\sum_{{\bf k}\neq0}\big|k^{-(c+\imath y)}\big|=
\sum_{{\bf k}\neq0}k^{-c}
\sim\int_1^{+\infty}\r^{d-1-c}\di\r<+\infty
$$
whenever $c>d$.
\begin{Lemma}
\label{resmainth}
The meromorphic function $\G(z)Z_d(z/2)\b^{-z}$ has all its poles on the real axis, for $z=d$, $z=0$ and for the
odd negative integers, whose residues are respectively:
\begin{itemize}
\item[(i)] ${\rm Res}_{z=d}\big[\G(z)Z_d(z/2)\b^{-z}\big]=L(\b)$;
\item[(ii)] ${\rm Res}_{z=0}\big[\G(z)Z_d(z/2)\b^{-z}\big]=-1$;
\item[(iii)] ${\rm Res}_{z=(1-2m)}\big[\G(z)Z_d(z/2)\b^{-z}\big]=C_m \b^{2m-1}$, with
\begin{equation}
\label{resdis}
C_m:=-\frac1{(2m-1)!}Z_d\bigg(\frac{1-2m}2\bigg)\,, \quad m=1,2,\dots\,\,\,.
\end{equation}
\end{itemize}
\end{Lemma}
\begin{proof}
We start noticing that, first 
the Epstein Z-function $Z_d(z)$  has only a simple pole at $z=d/2$ with residue $\frac{\pi^{d/2}}{\G(d/2)}$, see \cite{TerrasI}, Sec. 1.4, Thm. 1.
Second, the Euler Gamma function $\G(z)$ has only simple poles at $z=-n$, $n\in \bn$,  with $\text{Res}_{z=-n} =\frac{(-1)^n}{n!}$, see \cite{TerrasI}, Chap. 1.
Third, $Z_d(0)=-1$, and $Z_d(-m)=0$ for $m=1,2,3,\dots$, see \cite{TerrasI}, Sec. 1.4, Thm. 1. Finally, 
\begin{equation}\label{rela}
\begin{split}
Z_d\left(\frac{1-2m}{2}\right) &= \underbrace{\pi^{1-2m - d/2}}_{>0} \cdot \underbrace{\frac{\Gamma(\frac{d-1}{2} + m)}{\Gamma(\frac{1-2m}{2})}}_{\neq 0 \text{ and finite}} \cdot \underbrace{Z_d\left(\frac{d-1}{2} + m\right)}_{> 0}\\
&\neq 0 \text{ and finite}\ ,
\end{split}
\end{equation}
obtained by the {\it Reflection principle} of the Epstein function $Z_d(z)$
\begin{equation}\label{Reflec Z}
\pi^{-z} \Gamma(z) Z_d(z) = \pi^{-d/2 + z}\ \G\left(\frac{d}{2} - z\right) 
Z_d\left(\frac{d}{2} - z\right) ,
\end{equation}
see again \cite{TerrasI}, Sec. 1.4, Thm. 1.

\noindent
\textit{(i)} Since $Z_d(z)$ has a simple pole for $z=d/2$ with residue $\frac{\pi^{d/2}}{\G(d/2)}$, we get 
\[
\text{Res}_{z=d} Z_d\left(\frac{z}{2}\right) = \lim_{z \to d} (z-d) Z_d\left(\frac{z}{2}\right) = 2 \lim_{t \to d/2} \left(t - \frac{d}{2}\right) Z_d(t) = \frac{2\pi^{d/2}}{\G(d/2)}\ .
\]  
Now, both Euler Gamma function  $\G(d)$ and $\b^{-d}$ are regular and nonzero at $z=d$, and thus 
\[
{\rm Res}_{z=d}\big[\G(z)Z_d(z/2)\b^{-z}\big]
= \Gamma(d) \b^{-d}\text{Res}_{z=d} Z_d\left(\frac{z}{2}\right)
 =\frac{2\pi^{d/2}\Gamma(d)}{\G(d/2)\b^{d}} =L(\b)\ ,
\]
see \eqref{L beta}.

\noindent
\textit{(ii)} The Euler Gamma function $\G(z)$ has a simple pole at $z=0$ with residue $1$ and, since $Z_d(0)=-1$, we deduce
${\rm Res}_{z=0}\big[\G(z)Z_d(z/2)\b^{-z}\big]=-1$.

\noindent
\textit{(iii)} Taking account the above considerations and reasoning as before, for $n=1,2,\dots$ we deduce:
\begin{itemize}
\item[a)] if $n=2m$ is even, $\G(z)Z_d(z/2)\b^{-z}$ has a removable singularity at $z=-2m$ because $Z_d(-m)=0$;
\item[b)] if $n=2m-1$ is odd, $\G(z)Z_d(z/2)\b^{-z}$,  $\Gamma(z)$ has a simple pole at $z=1-2m$, as  $Z_d((1-2m)/2)\neq 0$ (e.g.\ \eqref{rela}), with residue
\end{itemize}
\begin{equation*}
\begin{split}
{\rm Res}_{z=1-2m}\big[\Gamma(z) Z_d(z/2)\b^{-z} \big]&=
Z_d((1-2m)/2)\b^{2m-1} {\rm Res}_{z=1-2m}\Gamma(z)\\
&=-\frac{Z_d((1-2m)/2)}{(2m-1)!}\b^{2m-1}
=C_m \b^{2m-1}\ .
\end{split}
\end{equation*}
\end{proof}
\begin{Lemma}
Let $a<0$ not be a pole of the function $\G(s)Z_d(s/2)$. Then,
\begin{equation}
\label{toty4}
\begin{split}
S(\b)=L(\b)+&\sum_{a<r<0}{\rm Res}_{s=r}\big[\G(s)Z_d(s/2)\b^{-s}\big]\\
+&\frac1{2\pi\imath}\int^{a+\imath\infty}_{a-\imath\infty}\G(s)\b^{-s}Z_d(s/2)\di s\,.
\end{split}
\end{equation}
\end{Lemma}
\begin{proof}
We apply the residue theorem to the rectangle $Q$ in the complex plane with boundary made of segments
$c+\iota[-Y,Y]$, $[c,a]+\iota Y$, $a+\iota[Y,-Y]$, $[a,c]-\iota Y$, $Y>0$, obtaining
\begin{equation*}
\begin{split}
\frac1{2\pi\imath}&\int_{\partial Q}\G(z)Z_d(z/2)\b^{-z}\di z={\rm Res}_{s=d}\big[\G(s)Z_d(s/2)\b^{-s}\big]\\
+&{\rm Res}_{s=0}\big[\G(s)Z_d(s/2)\b^{-s}\big]
+\sum_{a<r<0}{\rm Res}_{s=r}\big[\G(s)Z_d(s/2)\b^{-s}\big]\,.
\end{split}
\end{equation*}
We first note that 
${\rm Res}_{s=d}\big[\G(s)Z_d(s/2)\b^{-s}\big]=L(\b)$, and ${\rm Res}_{s=0}\big[\G(s)Z_d(s/2)\b^{-s}\big]=-1$ cancelling the constant term $1$ of $S(\b)$, see \eqref{toty}. The result now holds if one shows that
$$
\lim_{Y\to+\infty}\sup|\G(z)Z_d(z/2)\b^{-z}|=0\,,
$$
where the sup is taken for $z$ running on the horizontal segments. But this property follows from \eqref{toty2} and the fact that
$$
\sup_{a\leq\Re(z)\leq 2d}|Z_d(z/2)|=O\big(|\Im(z)|^{d-2a}\big)\,,
$$ 
which also guarantees that the integral in \eqref{toty4} is meaningful.
\end{proof}
Here is the asymptotic expansion of $S(\b)$ for $\b\downarrow0$.
\begin{Thm}
\label{mvtmh}
For the function $S(\b)$, we get
\begin{equation}
\label{toty3}
S(\b)=\bigg(\frac{2\pi^{d/2}\G(d)}{\G(d/2)}\bigg)\frac1{\beta^d}+\sum_{m=1}^{+\infty}C_m\b^{2m-1}\,,\quad0<\beta<2\pi\,,
\end{equation}
with $C_m$ given in \eqref{resdis}.
\end{Thm}
\begin{proof}
We start by noticing that the function in the integrand in \eqref{toty4} has negative poles in points $a_n=-n$, $n=1,2, \dots$. Therefore, we can choose for vertical line of the integral there as $a_n:=-(n+1/2)$. On the other hand, one easily gets
\begin{align*}
&\bigg|\frac1{2\pi\imath}\int^{-(n+1/2)+\imath\infty}_{-(n+1/2)-\imath\infty}\G(s)\b^{-s}Z_d(s/2)\di s\bigg|\\
\leq&\frac{\b^{(n+1/2)}}{2\pi}\int_{-\infty}^{+\infty}\big|\G(-(n+1/2)+\imath t)Z_d((-(n+1/2)+\imath t)/2)\big|\di t\\
=&K_n\b^{(n+1/2)}=o(\b^n)\,,
\end{align*}
where the last integral in the above formula is convergent as before.

By Proposition \ref{anal}, we know that $S(\b)-L(\b)$ extends to a holomorphic function in the open disk in the complex plane centred in $0$ of radius $2\pi$. On the other hand, 
the above computation tells us that the remainder in \eqref{toty4} is of order $o(\b^n)$.
By Peano theorem and Lemma \ref{resmainth}, we immediately conclude that the first sum there provides the Maclaurin expansion
of $S(\b)-L(\b)$ for all order $n\in\bn$, see the discussion at the end of Section \ref{pecoj}. 

Summarising, since the radius of convergence of the series in \eqref{toty3} is $2\pi$, for $0<\b<2\pi$ we obtain precisely the expansion we are searching for. 
\end{proof}
\begin{Rem}
As an immediate consequence of Proposition \ref{anal}, we get that the series in the r.h.s. of \eqref{toty3} has $2\pi$ as radius of convergence with respect to the complex variable $\b$.
\end{Rem}

\subsection{Bose and Fermi case for tori}
\label{32BF}

Such cases directly follow from Section \ref{1mic1}. Indeed, taking into account \eqref{occiiii} and reducing the matter to the case when the activity $z=1/2$, we have to handle
\begin{equation}
\label{notha}
\begin{split}
S_\pm(\b):=&\sum_{{\bf k}\in\bz^d}\frac1{2e^{\beta k}\pm1}
=\sum_{{\bf k}\in\bz^d}\Big(\sum_{l=1}^{+\infty}2^{-l}(\mp1)^{l+1}e^{-l\b k}\Big)\\
=&\sum_{l=1}^{+\infty}2^{-l}(\mp1)^{l+1}\Big(\sum_{{\bf k}\in\bz^d}e^{-l\b k}\Big)
=\sum_{l=1}^{+\infty}2^{-l}(\mp 1)^{l+1}S(l\b)
\,,\quad\beta>0\,,
\end{split}
\end{equation}
where $S$ is the function in \eqref{essebe} and $\pm$ correspond to the Fermi/Bose alternative. Notice that, for the Bose case, in \eqref{notha}
the sums can be freely exchanged, because the summand is positive, providing a finite value for each $\b>0$. Since
$$
|S_+(\b)|\leq\sum_{{\bf k},l}2^{-l}e^{-l\b k}=S_-(\b)\,,
$$
Lebesgue dominated convergence theorem allows the free exchange of both sums (in ${\bf k}$ and $l$) in \eqref{notha} also for the Fermi case.

Consider now the (analytic continuation of the) {\it polylogarithm} 
$$
{\rm Li}_s(z):=\sum_{l=1}^{+\infty}\frac{z^l}{l^s}\,,
$$
see \cite{Fla}, Sec. VI. 8.
\begin{Thm}
\label{nocox0}
The function in \eqref{notha} has the form
$$
S_\pm(\b)=\mp\bigg(\frac{2\pi^{d/2}\G(d){\rm Li}_{d}(\mp1/2)}{\G(d/2)}\bigg)\frac1{\beta^d}+R_\pm(\b)\,,
$$
where $R_\pm(\b)\in C^{\infty}(0)$ with Maclaurin expansion given by
\begin{equation}
\label{MLE}
R_\pm(\b)=\mp\sum_{m=1}^{n}C_m{\rm Li}_{1-2m}(\mp1/2)\b^{2m-1}+o\big(\beta^{2n})\,.
\end{equation}
\end{Thm}
\begin{proof}
Since the Fermi case is exactly similar, we reduce the matter to the Bose one (minus situation in \eqref{notha}), putting $f(\b):=S_-(\b)$. In this case,
$f(\b)=\La(\b)+{\rm P}(\b)$, where $\La(\b)$ is the (divergent) leading term and ${\rm P}(\b)$ is the finite remainder. By using \eqref{notha}, \eqref{essebe} and \eqref{L beta}, we easily obtain
\begin{align*}
\La(\b)=&\sum_{l=1}^{+\infty}2^{-l}L(l\b)=\bigg(\frac{2\pi^{d/2}\G(d)}{\G(d/2)}\bigg)\sum_{l=1}^{+\infty}\frac{2^{-l}}{(l\beta)^d}
=\bigg(\frac{2\pi^{d/2}\G(d)}{\G(d/2)}\bigg)\bigg(\sum_{l=1}^{+\infty}\frac{2^{-l}}{l^d}\bigg)\frac1{\beta^d}\\
=&\bigg(\frac{2\pi^{d/2}\G(d)}{\G(d/2)}\bigg)\frac{{\rm Li}_{d}(1/2)}{\beta^d}
=\bigg(\frac{2\pi^{d/2}\G(d){\rm Li}_{d}(1/2)}{\G(d/2)}\bigg)\frac1{\beta^d}\,.
\end{align*}

Concerning the remainder, we also use \eqref{R(b)} by noticing that it is meaningful for all $\b\in\br$.
Reasoning as before, apart from the coefficient and the factor $\b$ (both irrelevant) we have to check the differentiability of the remainder 
\begin{equation}
\label{propti}
{\rm P}(\b)\propto\sum_{l=1}^{+\infty}\sum_{{\bf k}\neq0}\frac{l2^{-l}}{(l^2\b^2+4\pi^2k^2)^{(d+1)/2}}\,,\quad \b\in\br\,.
\end{equation}
For the derivatives, with $|\b|\leq b$, we get
$$
\bigg|\frac{\di^n\,\,\,}{\di\b^n}\frac1{(l^2\b^2+4\pi^2k^2)^{(d+1)/2}}\bigg|\leq K\frac{l^{2n}}{k^{d+1}}\,,
$$
where $K:=K_{n,b}$ is a positive constant depending only on the order $n$ of the derivative and on the length of the interval on which $\b$ is running.
Therefore
\begin{align*}
\sum_{l=1}^{+\infty}\sum_{{\bf k}\neq0}\bigg|\frac{\di^n\,\,\,}{\di\b^n}\frac{l2^{-l}}{(l^2\b^2+4\pi^2k^2)^{(d+1)/2}}\bigg|
\leq&K\sum_{l=1}^{+\infty}l^{2n+1}2^{-l}\sum_{{\bf k}\neq0}\frac1{k^{d+1}}\\
\sim&\sum_{l=1}^{+\infty}\frac{l^{2n+1}}{2^{l}}\int_1^{+\infty}\frac{\di\r}{\r^2}<+\infty\,.
\end{align*}

The above computations tell us that the double series in the r.h.s.\ of \eqref{propti} can be freely differentiated  term by term infinitely often on all real line, deducing that
${\rm P}\in C^{\infty}(\br)$ and, a fortiori, that ${\rm P}$ is infinitely often differentiable in $0$.

Since ${\rm P}\in C^{\infty}(0)$, it admits the Maclaurin expansion of all orders. We then can easily end the proof in the following way. By the above consideration, 
${\rm P}(\b)=\sum_{l=1}^{+\infty}2^{-l}R(l\b)$ 
is $C^{\infty}(\br)$, and the above series can be differentiated term by term infinitely many times, obtaining
$$
{\rm P}^{(n)}(0)=\sum_{l=1}^{+\infty}2^{-l}l^n R^{(n)}(0)\,.
$$
We note that, if $n$ is even, then $R^{(n)}(0)=0$. If $n=2m-1$, we get
$$
{\rm P}^{(2m-1)}(0)={\rm Li}_{1-2m}(1/2)C_m(2m-1)!\,,
$$
which leads to \eqref{MLE} for the Bose case ${\rm P}(\b)=R_-(\b)$. 
\end{proof}
We now see that the remainder $R_\pm(\b)$ cannot extend to a holomorphic function in the neighbourhood of 0 in the complex plane.
\begin{Prop}
\label{00nocox00}
The series $\sum_{m=1}^{\infty}C_m{\rm Li}_{1-2m}(\mp1/2)\b^{2m-1}$ has zero radius of convergence and thus, according to
\eqref{mclau2}, we write
\begin{equation}
\label{nocox}
S_\pm(\b)\sim\mp\bigg(\frac{2\pi^{d/2}\G(d){\rm Li}_{d}(\mp1/2)}{\G(d/2)}\bigg)\frac1{\beta^d}
\mp\sum_{m=1}^{\infty}C_m{\rm Li}_{1-2m}(\mp1/2)\b^{2m-1}.
\end{equation}
\end{Prop}
\begin{proof}
Theorem \ref{nocox0} immediately leads to the 1st part, that is, apart from the leading term $\La_\pm (\b)$, the remainder ${\rm P}_\pm(\b)$ is infinitely often differentiable in $0$.
Concerning the 2nd part, 
that is the convergence of the series appearing in \eqref{nocox},
we first note that the
${\rm Li}_{1-2m}(\mp1/2)/(2m-1)!$ are the coefficients of the Maclaurin expansion of the functions $\frac1{2e^x\pm1}$ \cite[Thm. VI.7]{Fla}, obtaining, 
\begin{equation}
\label{nocoxx}
\bigg|\frac{{\rm Li}_{1-2m}(\mp1/2)}{(2m-1)!}\bigg|
\sim\begin{cases}
\frac1{\big(\sqrt{(\ln2)^2+\pi^2}\big)^{2m-1}}\,,\,\, -\,\,\text{(i.e. Fermi case)}\,,\\
\frac1{(\ln2)^{2m-1}}\,,\,\,+\,\,\text{(i.e. Bose case)}\,,
\end{cases}
\quad m\to\infty\,.
\end{equation}
On the other hand, by applying the reflection principle for $Z_d$, see \eqref{Reflec Z},
and by the $\G$-function reflection formula \cite[Eq. (2.1.20)]{PaKa}, we get
$\Gamma\left(\frac{1}{2} - m\right) \Gamma\left(\frac{1}{2} + m\right) = \pi(-1)^m$. Hence 
\[
Z_d\left(\frac{1-2m}{2}\right) = \frac{(-1)^m}{\pi^{2m+d/2}}\G\left(m + \frac{d-1}{2}\right) \G\left(m + \frac{1}{2}\right) Z_d\left(m + \frac{d-1}{2}\right) .
\]
Since, by the Lebesgue dominated convergence theorem,    $Z_d(m+(d-1)/2)\to 2d$ for $m\to+\infty$, we have to check the asymptotics of the two $\G$-functions. Using the Stirling formula, we get $\G(m+a)\sim m^{a-1}m!$ and thus, up to an inessential factor,
\begin{equation}
\label{nocoxxx}
\G(m+(d-1)/2)\, \G(m+1/2)\sim (m!)^2\,,\quad m\to\infty\,.
\end{equation}
Collecting \eqref{nocoxx} and \eqref{nocoxxx}, we deduce that the radius of convergence of the series in \eqref{nocox} is indeed zero.
\end{proof}

\subsection{Boltzmann case for the 3-sphere}
\label{tresf}

We now deal with the 3-sphere of radius $1$
$$
\bs^3:=\{(x,y,z,t)\in\br^4\,;\, x^2+y^2+z^2+t^2=1\}\,,
$$
since the 1-sphere coincides with the unit circle corresponding to the previous cases with $d=1$, the even-dimensional spheres are simpler (e.g. Section \ref{outshj}), and thus it represents the simplest nontrivial case.
The case of the other odd-spheres $\bs^{2l+1}\subset\br^{2(l+1)}$, $l>1$, can be similarly handled with technical and computational complications increasing with the dimension.
The spectrum of the Dirac operator of $\bs^3$ is 
$\{\pm(n+3/2)\,;\, n\in\bn\}$ with multiplicity $(n+1)(n+2)$, see e.g. \cite{G}, Theorem 2.1.3. 

Concerning the asymptotics of the number function, apart from inessential coefficients, we get
\begin{equation}
\label{esvarg1}
T(\b):=\frac14{\rm Tr}\big(e^{-\b|D|}\big)=\frac12\sum_{n=1}^{+\infty}n(n+1)e^{-\b(n+1/2)}=\frac1{8\sinh^3(\b/2)}\,,\quad\b>0\,,
\end{equation}
where the last equality is obtained differentiating once and twice with respect to
$\b$ under the sign of sum $\sum_n e^{-\b n}$, and then after standard algebraic manipulations.

\begin{Thm}
The function in \eqref{esvarg1} admits the asymptotic expansion
\begin{equation}
\label{esvarg11}
T(\b)=\frac1{\b^3}-\frac1{8\b}+\sum_{n=1}^{+\infty}\widetilde{C}_n\b^{2n-1}\ , \quad 0<\b<2\pi\ ,
\end{equation}
where, for $n=1,2,\dots$,
\begin{equation}
\label{esvarg10}
\widetilde{C}_n=\frac1{2^{2(n+1)}}\bigg[\frac{(1-2^{2n+1})B_{2(n+1)}}{2(n+1)(2n-1)!}-\frac{(1-2^{2n-1})B_{2n}}{(2n)!}\bigg]\,,
\end{equation}
$B_n$ being the Bernoulli numbers \cite[Ch. 24]{Frank}.
\end{Thm}
\begin{proof}
The proof easily follows by putting $x:=\b/2$ and looking at
$$
\frac{\b^3}{8\sinh^3(\b/2)}=\bigg(\frac{x}{\sinh x}\bigg)^3\,,
$$
noticing that the complex-valued function $\frac{z}{\sinh z}$ has a removable singularity in $z=0$, whose Maclaurin expansion can be found in \cite{GrRy, Frank}. Indeed,
\begin{align*}
\b^3 T(\b) =\left(\frac{\b/2}{\sinh(\b/2)} \right)^3 =1-\frac{\b^2}{8}+\sum_{n=1}^{+\infty}\widetilde{C}_n\b^{2(n+1)}\,, \quad 0<\b<2\pi\ ,
\end{align*}
and, by dividing by $\b^3$ both members, we obtain \eqref{esvarg11}.
\end{proof}

\subsection{Bose and Fermi case for the 3-sphere}

We now pass to the Bose and Fermi case for the $3$-dimensional sphere. For $\b>0$, we have
\begin{equation}
\label{esvarg3}
\begin{split}
T_\pm(\b):=&\frac14{\rm Tr}\Big[\big(2e^{\b|D|}\pm1\big)^{-1}\Big]
=\frac12\sum_{n=1}^{+\infty}n(n+1)\sum_{l=1}^{+\infty}2^{-l}(\mp1)^{l+1}e^{-l\b(n+1/2)}\\
=&\sum_{l=1}^{+\infty}2^{-l}(\mp1)^{l+1}\bigg(\frac12\sum_{n=1}^{+\infty}n(n+1)e^{-l\b(n+1/2)}\bigg)\\
=&\sum_{l=1}^{+\infty}2^{-l}(\mp1)^{l+1}T(l\b)
=\sum_{l=1}^{+\infty}2^{-l}(\mp1)^{l+1}\frac{1}{8\sinh^3[(l\b)/2]}\ .
\end{split}
\end{equation}
\begin{Thm}
The function in \eqref{esvarg3} has the form
$$
T_\pm(\b)=\mp{\rm Li}_{3}(\mp1/2)\frac1{\b^3}\pm\frac{{\rm Li}_{1}(\mp1/2)}8\frac1{\b}+\La_\pm(\b)\,,
$$
where $\La_\pm(\b)$ is $C^{\infty}$ in a neighbourhood of $\b=0$ with Maclaurin expansion given by
\begin{equation*}
\La_\pm(\b)=\mp\sum_{m=1}^{n}\widetilde{C}_m{\rm Li}_{1-2m}(\mp1/2)\b^{2m-1}+o\big(\beta^{2n})\,,
\end{equation*}
$\widetilde{C}_m$ given in \eqref{esvarg10}.
\end{Thm}
\begin{proof}
As before, we deal with the Bose case by putting $\Th(\beta):=T_-(\b)$ and $\D(\b):=\La_-(\b)$. Concerning the two divergent terms,
$$
\Th(\beta)-\D(\b)=\sum_{l=1}^{+\infty}\frac1{2^{l}(l\b)^3}-\frac18\sum_{l=1}^{+\infty}\frac1{2^{l}(l\b)}
=\frac{{\rm Li}_{3}(1/2)}{\b^3}-\frac{{\rm Li}_{1}(1/2)}{8\b}\,.
$$
Concerning the remainder, we consider $g(\b):=T(\b)-\frac1{\b^3}+\frac1{8\b}$, with $T(\b)$ given in \eqref{esvarg11}. It uniquely defines a smooth and bounded function on the real line with all derivatives again bounded. Therefore, the series defining the bounded part
$$
\D(\b)=\sum_{l=1}^{+\infty}2^{-l}g(l\b)
$$
can be differentiated term by term, providing the derivatives of $\D(\b)$. Hence, we argue that $\D\in C^\infty(\br)$. The assertion now follows as in Theorem \ref{nocox0}.
\end{proof}

Also in this case, we have
\begin{Prop}
\label{000nocox000}
The series $\sum_{m=1}^{+\infty}\widetilde{C}_m{\rm Li}_{1-2m}(\mp1/2)\b^{2m-1}$ has zero radius of convergence. Therefore, since
it cannot extend to a holomorphic function in the neighbourhood of 0 in the complex plane, according to
\eqref{mclau2} we write
\begin{equation*}
T_\pm(\b)\sim\mp{\rm Li}_{3}(\mp1/2)\frac1{\b^3}\pm\frac{{\rm Li}_{1}(\mp1/2)}8\frac1{\b}
\mp\sum_{m=1}^{+\infty}\widetilde{C}_m{\rm Li}_{1-2m}(\mp1/2)\b^{2m-1}\,.
\end{equation*}
\end{Prop}
\begin{proof}
The proof proceeds as in Proposition \ref{00nocox00}, by looking at the asymptotics of the sequence $(2m-1)!\widetilde{C}_m$. But, since the $\widetilde{C}_m$ are the coefficients of the Maclaurin series of (the extension to the whole real line of)
$$
T(\b)-\frac1{\b^3}+\frac1{8\b}=\frac1{8\sinh^3(\b/2)}-\frac1{\b^3}+\frac1{8\b}\,,
$$
we get 
\begin{equation}
\label{nocox0000}
\lim_m|\widetilde{C}_m|^\frac1{2m-1}=\frac1{2\pi}
\end{equation}

By using \eqref{nocoxx} and \eqref{nocox0000}, we get for the radius of convergence,
\begin{align*}
1/r=&\lim_m\big|\widetilde{C}_m{\rm Li}_{1-2m}(\mp1/2)\big|^{\frac1{2m-1}}
=\lim_m\bigg(\big|(2m-1)!\widetilde{C}_m\big|\bigg|
\frac{{\rm Li}_{1-2m}(\mp1/2)}{(2m-1)!}\bigg|\Bigg)^{\frac1{2m-1}}\\
=&\lim_m\big|(2m-1)!\widetilde{C}_m\big|^{\frac1{2m-1}}
\lim_m\bigg|\frac{{\rm Li}_{1-2m}(\mp1/2)}{(2m-1)!}\bigg|^{\frac1{2m-1}}=+\infty\,.
\end{align*}
\end{proof}

\section{Massive relativistic cases}
\label{mhdrcx}

The relativistic massive case is more complicated, and also more interesting. We see that a logarithmic term, containing $\beta\ln\beta$ as a factor, appears for odd dimensions. This seems confirmed also for the $d$-spheres $\bs^d\subset\br^{d+1}$ when $d$ is odd, even if all such models behave as a one-dimensional object (i.e.\ the spectrum is parametrised by a single parameter) from a computational point of view. Recall that, in this case,
\begin{equation}
\label{mrcz}
H(D)=c\sqrt{D^2+m^2c^2}\,,
\end{equation}
where $c$ is the speed of light and $m$ is the mass of the involved particles. To simplify, we put $c=1=m$. 

Define, 
$$
\z_d(z):=\sum_{{\bf k}\in\bz^d}\frac1{(k^2+1)^{z/2}}\,,\quad \Re(z)>d\,,
$$
and consider the meromorphic {\it shifted Epstein function} in this situation given by its analytic continuation, see e.g.\ \cite{EOR}, Ch. 2.

We also define
$$
\widetilde{\rm Li}_s(z):=\sum_{l=1}^{+\infty}\ln l\frac{z^l}{l^s}=-\frac{\partial{\rm Li}_s(z)}{\partial s}\,.
$$

\subsection{Boltzmann case for tori}

Let, for the dimension $d$,
$$
Q(\b):=\sum_{k\in{\mathbb Z}^d}e^{-\beta\sqrt{k^2+1}}=D(\b)+F(\b)\,,
$$
split into its divergent part $D(\b)$ and finite one $F(\b)$.
\begin{Thm}
\label{qudibeta}
For the function $Q(\b)=D(\b)+F(\b)$, we have for $0<\b<2\pi$,
\begin{equation}
\label{noperexaaa}
D(\b)= 2\pi^{d/2} \sum_{m=0}^{N} \frac{(-1)^m}{m!} \frac{\Gamma(d - 2m)}{\Gamma\left(\frac{d}{2} - m\right)} \beta^{-(d - 2m)}\,,
\quad N= \left\{
\begin{array}{ll}
d/2-1,\,\,&d\,\text{even}\,,\\
(d-1)/2,\,\,&d\,\text{odd}\,,\\
\end{array}
\right.
\end{equation}
\begin{equation}
\label{noperexa}
F(\b)=  \left\{
\begin{array}{ll}
\sum_{n=0}^\infty \frac{(-1)^n}{n!} \z_d(-n) \beta^n\,,\,&d\,\text{even}\,,\\
\sum_{j=0}^\infty \left[R_{2j+1} \ln \beta - A_{2j+1} - R_{2j+1} \psi(2j + 2) \right]\frac{\beta^{2j+1}}{(2j+1)!} \,\,&d\,\text{odd}\,,\\
\end{array}
\right.
\end{equation}
where
\begin{itemize}
\item $\psi(x) = \frac{\Gamma'(x)}{\Gamma(x)}=\frac{\di\ln\G(z)}{\di z}$ is the {\it digamma function}\,;
\item $R_{2j+1}$ is the residue of $\z_d(s)$ at the pole $s = -(2j+1)$:
$$
R_{2j+1} = \operatorname{Res}_{s = -(2j+1)} \z_d(s) = \frac{2(-1)^{\frac{d+1}{2}+j} \pi^{d/2}}{\left(\frac{d+1}{2} + j\right)! \Gamma\left(-j - \frac{1}{2}\right)}\,;
$$
\item $A_{2j+1}$ is the Hadamard finite part of $\z_d(s)$ in $s=-(2j+1)$:
$$
\lim_{s \to -(2j+1)}\bigg[\z_d(s) - \frac{R_{2j+1}}{s+2j+1}\bigg]\,.
$$
\end{itemize}
Both series in the variables $\b$ appearing in $F(\b)$ have radii of convergence $\geq2\pi$.
\end{Thm}
\begin{proof}
The proof follows the strategy of the case in Subsection \ref{1mic1}, though somewhat more involved. We outline the strategy and leave the details to the reader.

We first apply the Poisson summation formula to the function
$$
g(\bdx):= e^{-\beta\sqrt{x^2+1}}\,,\quad\bdx \in \mathbb{R}^d
$$
obtaining, after some tedious computations,
\begin{align*}
\widehat{g}(\bdxi) =&\frac{\pi^{\frac{d-1}{2}}\beta}{2} \int_0^\infty t^{-\frac{d+3}{2}} e^{-\left(t + \frac{\beta^2 + 4\pi^2\xi^2}{4t}\right)} \,\di t\\
=&2^{\frac{d+1}{2}} \pi^{\frac{d-1}{2}} \beta \left( \beta^2 + 4\pi^2\xi^2 \right)^{-\frac{d+1}{4}} K_{\frac{d+1}{2}}\Big(\sqrt{\beta^2 + 4\pi^2\xi^2}\Big)\, ,\quad\bdxi\in\br^d\,.
\end{align*}
Here $K_{\frac{d+1}{2}}(z)$ is the modified Bessel function of the second kind (see  \cite{wat}, or \cite[Ch. 10]{Frank}). Now, we first split $Q(\beta)$ in the zero-mode part and the nonzero-mode one:
$$
Q(\beta)=\widehat{g}({\bf 0})+\sum_{{\bf m}\neq0}\widehat{g}({\bf m})\,.
$$
For the nonzero-mode, since the argument of the involved Bessel function is $\sqrt{\beta^2 + 4\pi^2m^2}$ and the function $z^{-\nu} K_\nu(z)$ is analytic in $z^2$, each term $\widehat{g}(m)$ is a perfectly smooth, analytic function of $\beta^2$ in a neighbourhood of the origin with a radius of convergence restricted by the branch points at $\beta = \pm 2\pi\imath m$. Therefore, it contributes a regular power series whose radius of convergence is $2\pi$.

For the zero-mode,
$\widehat{g}({\bf 0}) = 2^{\frac{d+1}{2}} \pi^{\frac{d-1}{2}} \beta^{\frac{1-d}{2}} K_{\frac{d+1}{2}}(\beta)$ and, to obtain its contribution to the asymptotics of $Q(\b)$, we have to distinguish between the even and odd cases.

\medskip

\noindent
{\bf $d$ even}: in this case, the involved Bessel function is a polynomial in $1/\beta$, multiplied by $e^{-\beta}$. Its expansion for $\b\downarrow 0$ provides the divergent part (w.r.t.\ $\b$) in the expansion plus a convergent series summing up with that relative to the nonzero-mode.

\medskip

\noindent
{\bf $d$ odd}: in this case, the involved modified Bessel function is of the type $K_n(\beta)$, $n$ integer, and in this case it has a well-known logarithmic singularity at the origin, with expansion for $\b\approx 0^+$ is given by

\begin{align*}
K_n(\beta) =& \frac{1}{2} \left(\frac{2}{\beta}\right)^{n} \sum_{k=0}^{n-1} \frac{(n-k-1)!}{k!} (-1)^k \left(\frac{\beta}{2}\right)^{2k} + (-1)^{n+1} I_n(\beta) \ln\left( \frac{\beta}{2} \right)\\
+&\frac{(-1)^n}{2} \sum_{k=0}^\infty \frac{\psi(k+1) + \psi(n+k+1)}{k!(n+k)!} \left( \frac{\beta}{2} \right)^{2k+n}\,,
\end{align*}
where $I_n(\beta)$ is the modified Bessel function of the first kind.

Collecting together, we obtain the desired expansion of $Q(\b)$ for $\b\downarrow0$.
\end{proof}
The best way to compute \eqref{noperexaaa} and \eqref{noperexa} is to reason as in Subsection \ref{1mic1} by expressing $Q(\b)$ in terms of a Mellin anti-transform:
since
$$
\mathcal{M}\{Q\}(s) = \int_0^\infty \beta^{s-1} Q(\beta)\di\beta = \Gamma(s)\z_d(s)\,,
$$
it has the analytic continuation coinciding with that of the product of the meromorphic functions
$\Gamma(s)\z_d(s)$. Therefore
\begin{equation*}
Q(\b)=\frac1{2\pi\imath}\int^{c+\imath\infty}_{c-\imath\infty}\G(s)\z_d(s)\b^{-s}\di s\,,\quad c>d\,.
\end{equation*}
Therefore, in order to compute the asymptotics, we reason as in Theorem \ref{mvtmh} by looking at poles and the corresponding residues of the meromorphic function $I(s):=\Gamma(s) \zeta_d(s) \b^{-s}$.

To obtain the pole structure of the integrand function $\G(s)\z_d(s)\b^{-s}$, we must look at poles and zeroes of the functions $\Gamma(s)$ and $\zeta_d(s)$. Such an Epstein function is affected by the parity of the dimension $d$.

\medskip

\noindent
{\bf $d$ even}:
$\zeta_d(s)$ has simple poles at $s= d-2m$, for $m=0,1,2,\dots d/2-1$, 
and $\text{Res}_{s = d-2m} \zeta_d(s) = \frac{2 \pi^{d/2} (-1)^m}{m! \G \left(\frac{d}{2}-m\right)}$. This provides the finite sum in the top of \eqref{noperexaaa}. On the other hand, since the simple poles of the $\G$ are located in the nonpositive real axis $-\bn$ with residues $\frac{(-1)^n}{n!}$ and, correspondingly, $\z_d(s)\b^{-s}$ is analytic and nonzero there, we obtain the top of \eqref{noperexa}.

\medskip

\noindent
{\bf $d$ odd}: $\zeta_d(s)$ has simple poles at $s= d, d-2,\dots$.  Concerning those located in the positive part of the real axis, the residues have the same formula as $d$ even, obtaining the bottom part of \eqref{noperexaaa}. It remains to analyse what happens on the non positive real line because we have also the simple poles of the $\G$ there. 
For even nonpositive integers $s=-2j$, $j\in\bn$, the simple pole of the Gamma in $-2j$
``collides'' with the zero of the Epstein function, and thus $\G(s)\z_d(s)\b^{-s}$ has a removable singularity there. Hence, the nonpositive even integers do not contribute to the asymptotics of $Q(\b)$. If, instead, $s=-(2j+1)$, $j\in\bn$, the simple pole of $\G$ ``collides'' with the corresponding simple pole of $\zeta_d(s)$ providing a double pole. The corresponding residue of $\G(s)\z_d(s)\b^{-s}$ can be easily computed by the Laurent expansion of $\G(s)$ and $\z_d(s)$ and the Maclaurin expansion of $\b^{-s}$ around
$s_j=-(2j+1)$:
\begin{align*}
\G(s)=&\frac{\text{Res}_{s = s_j}\G(s)}{s-s_j}+\text{Pf}(\G)(s_j)+o(1)\,,\\
\zeta_d(s)=&\frac{\text{Res}_{s = s_j}\zeta_d(s)}{s-s_j}+\text{Pf}(\zeta_d)(s_j)+o(1)\,.
\end{align*}
Here the Pf denotes the {\it Hadamard finite part} of $f$ with a simple pole in $z_o$ given by
$$
\text{Pf}(f)(z_o):=\lim_{z\to z_o}\Big[\frac{\di\,\,}{\di z}\big((z-z_o)f(z)\big)\Big]=\lim_{z\to z_o}\bigg(f(z)-\frac{\text{Res}_{z =z_o}f(z)}{z-z_o}\bigg)\,.
$$
We easily obtain
\begin{align*}
&\text{Res}_{s = -(2j+1)}\big[\G(s)\zeta_d(s)\b^{-s}\big]=-\ln\b\text{Res}_{s = -(2j+1)}\G(s)\text{Res}_{s = -(2j+1)}\zeta_d(s)\b^{2j+1}\\
+&\text{Res}_{s = -(2j+1)}\G(s)\text{Pf}(\zeta_d)(-(2j+1))+\text{Pf}(\G)(-(2j+1))\text{Res}_{s = -(2j+1)}\zeta_d(s)\b^{2j+1}\,,
\end{align*}
recovering the bottom of \eqref{noperexa}.

\bigskip

We note that, in the odd case we find a logarithmic term of the form
$$
\b\ln\b\sum_{n=0}^{+\infty}b_n\b^{2n}\,,
$$
where the radius of convergence is strictly greater than 0 (indeed $+\infty$).
This means that, for this logarithmic term, 
$$
\lim_{\b\downarrow0}\bigg(\b\ln\b\sum_{n=0}^{+\infty}b_n\b^{2n}\bigg)
=b_0\lim_{\b\downarrow0}(\b\ln\b)=0\,.
$$
The appearance of this logarithmic term is clearly explained above with the Laurent expansion of $\G(s)\z_d(s)\b^{-s}$ around the double poles in the negative odd integers obtained by the ``collision'' of simple poles of $\G$ and $\z_d$.

\subsection{Bose and Fermi cases for tori}
\label{bfreft} 

For the Bose/Fermi relativistic massive cases, we reason as in Subsection \ref{32BF} as follows:
\begin{equation*}
\begin{split}
Q_\pm(\b):=&\sum_{k\in{\mathbb Z}^d}\frac1{2e^{\beta\sqrt{k^2+1}}\pm1}
=\sum_{{\bf k}\in\bz^d}\Big(\sum_{l=1}^{+\infty}2^{-l}(\mp1)^{l+1}e^{-l\b\sqrt{k^2+1}}\Big)\\
=&\sum_{l=1}^{+\infty}2^{-l}(\mp)^{l+1}Q(l\b)\,,\quad\beta>0\,.
\end{split}
\end{equation*}
Concerning $Q(\b)$, by Theorem \ref{qudibeta}, we can write
\[
Q(\b)=  \left\{
\begin{array}{ll}
\sum_{m=0}^{d/2-1} q_m\b^{-(d-2m)}
+\sum_{m=0}^\infty a_m\beta^m\,,\,&d\,\text{even}\,,\\
\sum_{m=0}^{(d-1)/2} q_m\b^{-(d-2m)}
+\b\ln\b\sum_{m=0}^\infty b_m\b^{2m}+\sum_{m=0}^\infty c_m\beta^{2m+1} \,\,&d\,\text{odd}\,,\\
\end{array}
\right.
\]
where the coefficients $q_m,a_m,b_m,c_m$ depend on the involved dimension $d=1,2,\dots\,\,\,$. Notice that the series appearing for even dimension and the 2nd series in the odd alternative have nonzero radius of convergence.

In this situation, we obtain
\begin{equation}
\label{fghxz}
Q_\pm(\b)\sim\mp\left\{
\begin{array}{ll}
\sum_{m=0}^{d/2-1} {\rm Li}_{d-2m}(\mp1/2)q_m\b^{-(d-2m)}
+\sum_{m=0}^\infty {\rm Li}_{-m}(\mp1/2)a_m\beta^m\,,\,&d\,\text{even}\,,\\
\sum_{m=0}^{(d-1)/2} {\rm Li}_{d-2m}(\mp1/2)q_m\b^{-(d-2m)}
+\sum_{m=0}^\infty \S_m(\b)\beta^{2m+1} \,\,&d\,\text{odd}\,,\\
\end{array}
\right.
\end{equation}
where
$$
\S_m(\b):=\Big(b_m\big(\widetilde{\rm Li}_{-2m-1}(\mp1/2)
+\ln\b{\rm Li}_{-2m-1}(\mp1/2)\big)+c_m{\rm Li}_{-2m-1}(\mp1/2)\Big)\,.
$$

Since the radii of convergence of $\sum_{m=0}^\infty a_m\beta^m$ and $\sum_{m=0}^\infty c_m\beta^{2m+1}$ are nonzero but finite, the resulting series in \eqref{fghxz} cannot have
finite radii of convergence.\footnote{Compare with Propositions \ref{00nocox00} and \ref{000nocox000}.} Therefore, as in all previous situations for $C^\infty(0)$ but not analytic functions, it is meant as finite truncations at each fixed order $n$ with remainder of order $o(\b^n)$. 

\subsection{The 3-sphere}

Here we describe what happens for the relativistic massive situation for the 3-sphere, being the simplest nontrivial case among spheres. In fact, the 1-sphere obviously coincides with the unit circle treated in Theorem \ref{qudibeta} (and Subsection \ref{bfreft} for the Bose and Fermi cases) when $d=1$, whereas the simpler cases of even-spheres are outlined in Section \ref{outshj}.
We see that, also in this situation, a logarithmic term 
containing $\b\ln\b$ appears in the asymptotic expansion.
We first deal with the Boltzmann case, and then recover the Bose/Fermi cases as before. 

Taking into account the spectrum, with the corresponding multiplicity, of the Dirac operator $D_{\bs^3}$ on the $\bs^3$ given in Subsection \ref{tresf} and, finally, disregarding an inessential multiplying factor, we put
$$
{\rm Tr}\Big(e^{-\b\sqrt{D^2_{\bs^3}+1}}\Big)\propto Q^{\bs^3}(\beta):=\sum_{k=1}^{+\infty}k(k+1)e^{-\beta\sqrt{(k+1/2)^2+1}}\,.
$$
\begin{Thm}
\label{qudibeta11}
For $0<\b<2\pi$, we have
\begin{equation}
\label{bestway}
Q^{\bs^3}(\beta)=\frac{2}{\beta^3} - \frac{3}{4\beta} + \sum_{j=0}^{+\infty} \left[ L_j \ln\left(\frac{\beta}{2}\right) + I_j + S_j \right] \beta^{2j+1}
\end{equation}
where
\begin{itemize}
\item $L_j = - \frac{j+4}{2^{2j+3} j! (j+2)!}$\,;
\item $I_0 = \frac{5-8\gamma}{32}$ and, for $j \geq1$, $I_j = \frac{2(j+2)[\psi(j+1)+\psi(j+2)] - j[\psi(j)+\psi(j+3)]}{2^{2j+4} j! (j+2)!}$;
\item $S_j = \frac{(-1)^{j+1}}{2^{j} j! (2\pi)^{j+1}} \sum_{k=1}^{+\infty} \frac{(-1)^k}{k^{j+1}} \left[ \frac{2j+3}{2\pi k} K_{j+2}(2\pi k) + \frac{5}{4} K_{j+1}(2\pi k) \right]$.
\end{itemize}
Here $\g$ is the Euler-Mascheroni constant, $\psi(z)$ is the digamma function, and $K_\n$ are modified Bessel functions, see  \cite{wat} or \cite[Ch. 10]{Frank}.

Both series in the variables $\b$ appearing in the asymptotic expansion of $Q^{\bs^3}(\beta)$
have radii of convergence $>0$.
\end{Thm}
\begin{proof}
The proof follows the lines of Theorems \ref{mvtmh} and \ref{qudibeta}. We sketch it and leave the details to the reader.

After a change of variables $n = k + 1/2$, and applying the Poisson summation formula, we get 
\begin{equation}
\label{edomczx}
Q^{\bs^3}(\beta)=\frac{1}{2} \sum_{n \in \mathbb{Z} + 1/2} \left(n^2 - \frac{1}{4}\right) e^{-\beta\sqrt{n^2+1}} = \frac{1}{2} \sum_{m=-\infty}^{\infty} (-1)^m \widehat{h}(m)\,,
\end{equation}
with $h(x) = (x^2 - 1/4) g(x)$ and $g(x) = e^{-\beta\sqrt{x^2+1}}$. Now, with $z = \sqrt{\beta^2 + (2\pi \xi)^2}$, we obtain the {\it Bessel Identity}
$$
\widehat{g}(\xi) = \int_{-\infty}^{\infty} e^{-\beta\sqrt{x^2+1}} e^{-2\pi i \xi x}\di x = 2\beta \frac{K_1(z)}{z}\,,\quad \xi\in\br\,.
$$
After some standard computations, we get
\begin{equation}
\label{edomczx1}
\widehat{h}(\xi) = 2\beta \left[ \frac{K_2(z)}{z^2} - (2\pi \xi)^2 \frac{K_3(z)}{z^3} - \frac{1}{4} \frac{K_1(z)}{z} \right]\,.
\end{equation}
Also in this situation, we split the computation in the zero and nonzero-mode in \eqref{edomczx}, taking into account \eqref{edomczx1}.

\medskip

\noindent
{\bf Zero-mode $m=0$}: Since at the discrete values $\xi=m$, $m=0$ corresponds $z=\b$, expanding the Bessel functions around $\b\approx0^+$ and regrouping all contributions (recall that the involved series have radii of convergence $>0$), we get the singular terms $\frac{2}{\beta^3} - \frac{3}{4\beta}$ plus
$\sum_{j=0}^{+\infty} \left[ L_j \ln\left(\frac{\beta}{2}\right) + I_j \right] \beta^{2j+1}$.

\medskip

\noindent
{\bf Nonzero-mode $m\neq0$}: Since for $\xi = m \neq 0$, $z = \sqrt{\beta^2 + (2\pi m)^2}>0$, 
$K_\nu(z)$ is analytic in $\beta^2$. We expand $z^{-\nu}K_\nu(z)$ as a Taylor series in $z = \beta^2$ around $z=0$. Each derivative brings down a factor of $-1/2$:
$$
\left(\frac{d}{dx}\right)^j \left[ \frac{K_\nu(z)}{z^\nu} \right] = \frac{(-1)^j}{2^j} \frac{K_{\nu+j}(z)}{z^{\nu+j}}\,.
$$
Taking that into account, expanding $\widehat{h}(m)$ into odd powers of $\beta$ via this operator, and utilising the Bessel recurrence relation 
$$
K_{\nu+1}(z) - K_{\nu-1}(z) = \frac{2\nu}{z}K_\nu(z)
$$ to simplify the degree of the polynomials, we provide exactly
$\sum_{j=0}^{+\infty} S_j  \beta^{2j+1}$.
\end{proof}
As before, the best way to compute the expansion in \eqref{bestway} is to reason as in Subsection \ref{1mic1} by expressing $Q(\b)$ in terms of a Mellin anti-transform. Indeed, for the Mellin transform,
$$
\mathcal{M}\big\{Q^{\bs^3}\big\}(s) = \int_0^\infty \beta^{s-1} Q^{\bs^3}(\beta)\di\beta = \Gamma(s) Z^{\bs^3}(s)
$$
(for values of $s\in\bc$ when the above integral is meaningful),
where $Z^{\bs^3}$ is the analytic continuation of the corresponding Epstein function given by
$$
Z^{\bs^3}(s) = \sum_{k=1}^{\infty} \frac{(k+1/2)^2 - 1/4}{((k+1/2)^2+1)^{s/2}}\,.
$$
This means that
\begin{equation*}
Q^{\bs^3}(\b)=\frac1{2\pi\imath}\int^{c+\imath\infty}_{c-\imath\infty}\G(s)\b^{-s}Z^{\bs^3}\!(s)\di s\,,\quad c>3\,.
\end{equation*}
Therefore, the expansion for $Q^{\bs^3}(\b)$ can be obtained by summing up all residues corresponding to the poles of the meromorphic function $\G(s)\b^{-s}Z^{\bs^3}\!(s)$, all located on the real line for $s=\Re(s)\leq3$.

By using the Poisson summation formula, we express the Epstein function in a more convenient way by using the standard Jacobi Theta-Null function identity 
$$
\sum_{k \in \mathbb{Z}} e^{-(k+1/2)^2 t} = \sqrt{\frac{\pi}{t}} \sum_{m \in \mathbb{Z}} (-1)^m e^{-\pi^2 m^2 / t}\,,
$$
obtaining, after some tedious computations,
\begin{equation}
\label{tt000011}
\begin{split}
\frac{\sqrt{\pi}}{\Gamma(s/2)} \int_0^\infty \!\di t\,t^{s/2-1} e^{-t} 
&\sum_{m=-\infty}^{\infty} (-1)^m \left( \frac{1}{4} t^{-3/2} - \frac{\pi^2 m^2}2 t^{-5/2} - \frac{1}{8} t^{-1/2} \right) e^{-\pi^2 m^2 / t}\\
=&Z^{\bs^3}\!(s) =\sum_{m=-\infty}^{\infty}Z_m^{\bs^3}\!(s)\,,
\end{split}
\end{equation}
after the exchange between the integral and the sum.

\medskip

\noindent
{\bf Poles from the zero-mode $m=0$}: For $m=0$ in \eqref{tt000011}, we have
$$
\Gamma(s)Z_0^{\bs^3}\!(s) = 2^{s-3} \Gamma\left(\frac{s+1}{2}\right) \Gamma\left(\frac{s-3}{2}\right) - 2^{s-4} \Gamma\left(\frac{s+1}{2}\right) \Gamma\left(\frac{s-1}{2}\right)\,.
$$
For $s=3$ and $s=1$ we have single poles with residues $2\beta^{-3}$ and $-\frac{3}{4}\beta^{-1}$, providing the divergent part of the expansion. If, instead,
$s_k=-(2k+1)$, $k\in\bn$ we have a double pole of $f(s)=\Gamma(s)Z_0^{\bs^3}\!(s)$ due to the ``collision'' of single poles of the Gammas. The corresponding residues can be easily computed as before by the Laurent series of $f(s)$ and the Maclaurin expansion of $\b^{-s}$:
\begin{align*}
&f(s)=\frac{\z_k}{(s-s_k)^2}+\frac{\eta_k}{(s-s_k)}+o(1)\,,\\
&\b^{-s}=\b^{-s_k}+(-\ln\b)\b^{-s_k}(s-s_k)+o(s-s_k)\,.
\end{align*}
Therefore, the corresponding residue is
$$
{\rm Res}_{s=-(2k+1)}\big[\Gamma(s)Z_0^{\bs^3}\!(s)\b^{-s}\big]=\big(-\z_k\ln\b+\eta_k\big)\b^{2k+1}=(L_k\ln(\b/2)+I_k)\b^{2k+1}\,.
$$
As before, this clearly explains why the logarithmic piece containing $\ln\b$ appears in \eqref{bestway}. Notice that no pole appears for negative even integers due to the ``collision'' of a simple pole of the Gamma function and a simple zero of a shifted Gamma function.

\medskip

\noindent
{\bf Poles from the nonzero-modes $m\neq0$}: The integrals are recognised immediately as the integral representation of the Bessel $K$-functions:
$$
\int_0^\infty t^{\nu-1} e^{-t - B/t} dt = 2B^{\nu/2}K_\nu(2\sqrt{B})\,.
$$
As before, due to a cancellation of the possible poles for even negative integers by the ``collision'' with a single zero, only single poles in negative odd integers will survive.
Evaluating the residues of $\Gamma(s)Z_m^{\bs^3}\!(s)\b^{-s}$ of these simple poles at $s = -(2k+1)$ (one for each summand) algebraically mirrors the coefficient extraction from the Poisson summation formula Taylor expansion, generating the exact same contribution $\sum_{j=0}^{+\infty} S_j  \beta^{2j+1}$ as before.

\bigskip

Now, some comments are in order as for the relativistic Boltzmann case. 
First we note that
\begin{equation}
\label{fghxzaa}
Q^{\bs^3}(\beta)=\frac{2}{\beta^3} - \frac{3}{4\beta}+\b\ln\b\sum_{m=0}^\infty b^{\bs^3}_m\b^{2m}+\sum_{m=0}^\infty c^{\bs^3}_m\beta^{2m+1}\,,
\end{equation}
where the radius of convergence of the 1st series is $+\infty$, and that of 2nd one is $2\pi$. Second, as before, for the logarithmic term we get
$$
\lim_{\b\downarrow0}\b\ln\b\sum_{m=0}^\infty b^{\bs^3}_m\b^{2m}=0\,.
$$

Finally, we can immediately handle the Bose and Fermi cases by using \eqref{fghxzaa}. We obtain
\begin{equation}
\label{spfunct1}
Q_\pm^{\bs^3}(\beta):=\sum_{k=1}^{+\infty}\frac{k(k+1)}{2e^{\beta\sqrt{(k+1/2)^2+1}}\pm1}
\sim\mp\frac{2{\rm Li}_{3}(\mp1/2)}{\b^{3}}\pm\frac{3{\rm Li}_{1}(\mp1/2)}{4\b}
\mp\sum_{m=0}^\infty \S^{\bs^3}_m(\b)\beta^{2m+1}\,,
\end{equation}
where
$$
\S^{\bs^3}_m(\b):=\Big(b^{\bs^3}_m\big(\widetilde{\rm Li}_{-2m-1}(\mp1/2)
+\ln\b{\rm Li}_{-2m-1}(\mp1/2)\big)+c^{\bs^3}_m{\rm Li}_{-2m-1}(\mp1/2)\Big)\,.
$$

\section{Outlook}
\label{outshj}

In the present paper, which is a natural continuation of the previous one \cite{CiF}, we provide the asymptotics of the spectral actions associated with the ``number function'' of theories describing noninteracting particles. 
To the best of the authors' knowledge,
the model consisting of a gas of free particles in thermal equilibrium seems to be the only realistic one, 
allowing such kind of concrete computations which,
of course, might have concrete physical applications.
After a long list of works dealing with such an asymptotic analysis and its applications also in number theory 
(e.g. \cite{EI, EGV}), this investigation follows the initial attempt in \cite{DKS} to investigate the spectral action arising from models describing free particles. Such an investigation is relevant, not only for potential physical applications, but also for other aspects involving the ``geometry'' of the underlying manifold (e.g. 
\cite{CC, CC1, EI}  and the literature cited therein).
The relevant merit of this model based on free particles is that it is possible to carry out explicit, even if very complicated, computations.

The first aspect we would like to point out is the appearance of divergent power series (i.e.\ when the radius of convergence is 0). It happens in all situations, treated here, and relative to Bose and Fermi cases where the involved partition functions are merely of class $C^\infty(0)$. Therefore, the other connected interesting aspect concerns the possible appearance of a remainder of order $o(\beta^{\infty})$ for $\b\downarrow0$ which cannot be detected when the involved function is only infinitely often differentiable in 0.
To provide a precise definition according which such an infinitesimal term appears or not is a very delicate question of asymptotic analysis treated in the specific literature (i.e. \cite{EK, Cos}). A heuristic criterion to decide whether such a term indeed appears is explained in Section \ref{pecoj} and involves (generalisations of) the Borel summation process of divergent series. In the present situations, after following this criterion (but without providing a possibly long and involved proof), one might infer that such an infinitesimal term never appears.

Another very interesting aspect is the appearance of the logarithmic term containing $\b\ln\b$ as a factor in the massive relativistic case, when the dimension is odd (cf. \eqref{noperexa}) and in the 3-sphere (cf.
\eqref{spfunct1}). Since the 1-sphere coincides with the case in \eqref{noperexa} when $d=1$, in order to understand what is happening, we outline the cases of even-spheres.

Indeed, for the $2n$ sphere $\bs^{2n}$, consider the relative {\it Epstein functions} as the analytic continuation (in the $s$-variable) of
$$
D_n(s):=\sum_{k=1}^{+\infty}\frac{k(k+1)\cdots(k+2(n-1))}{\big((k+n-1)^2+1\big)^{s/2}}\,.
$$
Now, by using the same strategy as before connected with the Mellin (anti-)transform, we can deduce that, 
$$
{\rm Tr}\Big(e^{-\b\sqrt{D^2_{\bs^{2n}}+1}}\Big)\propto\sum_{k=0}^{n-1}\frac{d_k^{(n)}}{\b^{2(n-k)}}+\sum_{k=0}^{+\infty}\frac{(-1)^k D_n(-k)}{k!}\b^{k}\,,\quad\b>0\,,
$$
and the series in the r.h.s. above has $2\pi$ as radius of convergence. Therefore, for the even dimensional spheres, the logarithmic term containing $\b\ln\b$ does not appear.

The last case of interest, which deserves a deep investigation, is that relative to the model arising from the Dirac operator, deformed by using the Tomita modular operator and described in Section 7 of \cite{FS}. Such models
can play a role in constructing new noncommutative geometries involving type III $W^*$-algebras 
setting by emphasising the need to take the Tomita modular data into account. In this sense,
this framework would be the first ``genuine'' (and highly nontrivial) model of noncommutative geometry in which such computations might be carried out.
The problem to recover the asymptotics of the spectral action of the number function associated with such a deformed Dirac operator appears extremely difficult because the undeformed Dirac operator on the 2-torus and the Tomita modular operator do not commute with each other, and thus cannot be simultaneously diagonalised. The question to recover the spectrum of such a deformed Dirac operator,  treated in Section 8 of \cite{CF0}, is formulated as an eigenvalue problem for a class of Hill equations.

\section*{AI statements} The authors addressed AI free version of Gemini (google) which supported them in obtaining some relevant parts of their analysis, including the relative bibliography. All outputs of Gemini used in the present paper were carefully checked by the authors, who 
assume the full responsibility for the obtained results.

\section*{Acknowledgements}
F. Fidaleo acknowledges ``Excellence Department Project'' by Italian Government, CUP E83C23000330006; and ``Tor Vergata University of Rome funding OANGQS'', CUP E83C25000580005. F. Ciolli acknowledges the support of Italian INdAM-GNAMPA. The authors are grateful to the referee for highlighting some inaccuracies and for suggestions that contributed to improving the presentation of this paper.

\end{document}